\documentclass[final,5p,times]{elsarticle}
\usepackage{graphicx}
\usepackage{amsmath}
\usepackage{amsfonts}
\newcommand{\D}{{\mathrm d}}
\newcommand{\I}{{\mathrm i}}

\newcounter{bla}

\journal{Computer Physics Communications}

\begin{document}
\begin{frontmatter}

\title{The TDHF Code Sky3D}

\author{J.~A.~Maruhn\corref{author}} \ead{maruhn@th.physik.uni-frankfurt.de}
\address{Institut f\"ur Theoretische Physik, Goethe-Universit\"at,
  Max-von-Laue-Str. 1, \\60438 Frankfurt am Main, Germany}

\author{P.-G.~Reinhard}\ead{Paul-Gerhard.Reinhard@physik.uni-erlangen.de}
\address{Institut f\"ur Theoretische Physik II, Universit\"at
  Erlangen-N\"urnberg, \\Staudtstrasse 7, 91058 Erlangen, Germany}

\author{P.~D.~Stevenson}\ead{p.stevenson@surrey.ac.uk}
\address{Department of Physics, University of Surrey, Guildford,
  Surrey, GU2 7XH, United Kingdom}

\author{A. S. Umar}\ead{sait.a.umar@Vanderbilt.Edu}
\address{Department of Physics and Astronomy, Vanderbilt University, \\Nashville, Tennessee 37235, USA}

\cortext[author]{Corresponding author. \textit{Phone: +49-69-79847873}}

\begin{abstract}
The nuclear mean-field model based on Skyrme forces or related density
functionals has found wide-spread application to the description of
nuclear ground states, collective vibrational excitations, and
heavy-ion collisions. The code Sky3D solves the static or dynamic
equations in a three-dimensional Cartesian mesh with isolated or
periodic boundary conditions and no further symmetry
assumptions. Pairing can be included in the BCS approximation. The
code is implemented with a view to allow easy modifications for including
additional physics or special analysis of the results.
\end{abstract}
\end{frontmatter}

{\bf PROGRAM SUMMARY}

\begin{small}
\noindent
{\em Title:}  The TDHF Code Sky3D \\
{\em Authors:}   J.~A.~Maruhn, P.-G.~Reinhard, P.~D.~Stevenson, and A.~S.~Umar.\\
{\em Program Title:} Sky3D\\
{\em Journal Reference:}                                      \\
{\em Catalogue identifier:}                                   \\
{\em Licensing provisions:}   none\\
{\em Programming language:}  Fortran 90\\
{\em Computer:}   All computers with a Fortran compiler supporting at
least Fortran 90.\\
{\em Operating system:}  All operating systems with such a
compiler. Some of the Makefiles and scripts depend on a Unix-like
system and need modification under Windows.\\
{\em RAM:} 1~GB\\
{\em Number of processors used:} no built-in limit, runs under both
OpenMP and MPI
{\em Classification:}  17.16 Theoretical Methods - General, 17.22 
Hartree-Fock Calculations, 17.23 Fission and Fusion Processes\\
{\em External routines/libraries:}  LAPACK, FFTW3\\
\\
{\em Nature of problem:} The time-dependent Hartree-Fock equations can
be used to simulate nuclear vibrations and collisions between nuclei
for low energies. This code implements the equations based on a Skyrme
energy functional and also allows the determination of the
ground-state structure of nuclei through the static version of the
equations. For the case of vibrations the principal aim is to
calculate the excitation spectra by Fourier-analyzing the time
dependence of suitable observables. In collisions, the formation of a
neck between nuclei, the dissipation of energy from collective motion,
processes like charge transfer and the approach to fusion are of
principal interest.\\
\\
{\em Solution method:} The nucleonic wave function spinors are
represented on a three-dimensional Cartesian mesh with no further
symmetry restrictions. The boundary conditions are always periodic for
the wave functions, while the Coulomb potential can also be calculated
for an isolated charge distribution. All spatial derivatives are
evaluated using the finite Fourier transform method. The code solves
the static Hartree-Fock equations with a damped gradient iteration
method and the time-dependent Hartree-Fock equations with an expansion
of the time-development operator. Any number of initial nuclei can be
placed into the mesh in with arbitrary positions and
initial velocities.\\
\\
{\em Restrictions:} The reliability of the mean-field approximation is
limited by the absence of hard nucleon-nucleon collisions. This limits
the scope of applications to collision energies about a few MeV per
nucleon above the Coulomb barrier and to relatively short interaction
times. Similarly, some of the missing time-odd terms in the implementation
of the Skyrme interaction may restrict the applications to even-even
nuclei.\\
\\
   \\
{\em Unusual features:}\\
The possibility of periodic boundary conditions and the highly
flexible initialization make the code also suitable for astrophysical
nuclear-matter applications.\\
   \\
{\em Running time:} The running time depends strongly on the size of
the grid, the number of nucleons, and the duration of the
collision. For a single-processor PC-type computer it can vary between
a few minutes and weeks.\\
   \\
\end{small}

\tableofcontents

\section{Introduction}
The vast majority of microscopic models of many-body systems rely
on a description in terms of the single-particle (s.p.)
wave functions. Among them, self-consistent mean-field models (SCMF)
automatically generate the optimal one-body potentials
corresponding to the s.p. wave functions. A rigorous SCMF is the
Hartree-Fock theory (HF) where the s.p. wave functions are determined
variationally for a given two-body interaction~\cite{Fet71,Mar10aB}. A
more practical approach is provided by the Density Functional Theory (DFT),
which incorporates the involved many-body effects into effective
interactions, or effective energy-density functionals.
This is a very efficient and successful scheme, widely used in
electronic systems~\cite{Dre90}. Straightforward HF is unsuitable
for nuclei because the free-space two-nucleon force contains a strong
short-range 
repulsion requiring renormalization in the nuclear medium. 
For this reason, nuclear SCMFs necessarily employ effective
interactions or functionals although they often carry the label
HF as, e.g, in the Skyrme Hartree-Fock (SHF) method. There are
relativistic as well as non-relativistic approaches.  For an extensive
review see~\cite{Ben03aR}.

The description of dynamical processes is even more demanding than the
modeling of structure. SCMFs are the also the first method of choice
in this domain. The natural extension of HF is time-dependent HF
(TDHF) which was proposed as early as 1930 in~\cite{Dir30}. Earlier
applications were restricted to the linearized regime covering small
amplitude motion, see, e.~g., \cite{Bro71aB}. Large scale TDHF
calculations became possible in the last few decades with the
increasing computing capacities. Again, as in the static case, true
TDHF calculations make sense only for electronic systems and even
there they are very rare. The overwhelming majority of dynamical SCMF
calculations employ, in fact, time-dependent DFT (TDDFT). In
electronic systems, this amounts to the time-dependent
local density approximation (TDLDA) \cite{Dre90}, which is widely used
in atoms, molecules, and solids; for examples in nanoparticles see,
e.~g., \cite{Rei03a}. Dynamical SCMFs in nuclei also stay at the level
of TDLDA even if they are often named TDHF which happens particularly
for dynamical calculations using the Skyrme energy functional. Nuclear
TDHF started about forty years ago \cite{Bon76a} and has developed
since then into a powerful and versatile tool for simulating a great
variety of dynamical scenarios.  Earlier applications were based
mainly on non-relativistic TDHF using the effective Skyrme energy
functional~\cite{Neg82aR,Dav85a}.  Due to higher numerical demands,
relativistic calculations appeared somewhat later~\cite{Bai87a}, but
have developed meanwhile equally well to a widely used
tool~\cite{Ber95a,Vre05aR}.

In this paper, we present a code for TDHF calculations on the basis of
the non-relativistic Skyrme energy functional. The code uses a fully
three dimensional (3D) representation of wave functions and fields on
a Cartesian grid in coordinate space. There are no symmetry
restrictions and the full Skyrme energy functional is used including
the spin-orbit and most important time-odd terms. Such fully-fledged
3D calculations became possible only over the last decade with the
steadily increasing computing capabilities. In fact, early TDHF
studies all used restricted representations, axial symmetry and/or
reflection symmetries. This limited the possible applications. TDHF
experienced a revival during the last ten years when unrestricted 3D
calculations became possible.  There are several groups performing
large scale TDHF studies for various scenarios of nuclear dynamics,
see, e.g.,
\cite{Sim03a,Mar05a,Umar05a,Nak05a,Has08a,Uma09a,Loe12a}. 
For a recent review see \cite{Simenel}. Such
calculations have clearly outgrown the developmental stage. It is an
appropriate time to give a broader public access to a 3D TDHF
code. This is the goal of this paper. Skyrme HF covers such a broad
range of physical phenomena and is relatively involved that efficient
computational treatment of 3D simulations requires elaborate numerical
methods. We shall make an effort to explain the many necessary
ingredients in a comprehensive, and yet compact, manner.

Most recent Skyrme density functionals contain terms such as
fractional powers of the density that cannot be related to a two- or
three-body interaction. In that sense, the present code solves the
TDDFT rather than TDHF equations. Nevertheless, we prefer to keep the
name TDHF since it is associated historically with this large field of
nuclear reaction theory.

\section{General purpose and structure}

\subsection{Intended applications}
The code Sky3D solves the static Hartree-Fock as well as the
time-dependent Hartree-Fock (TDHF) equations for interactions of
Skyrme-force type in a general three-dimensional geometry. No
symmetries of any kind are assumed, so that the code can be used for a
wide variety of applications in nuclear structure, collective
excitations, and nuclear reactions; of course within the limitations
of mean-field theory.

\subsection{Specific model implemented}
The code in the presented version contains a useful selection of terms
in the Skyrme force but by no means all terms that have been included
in some recent works. It should still be useful, because (1) for many
interesting applications the interest is semi-quantitative so that a
Skyrme force fitted with the latest models is not necessary --- usually
a selection of forces is desired to look at the variability of
results, but not a high-accuracy fit of data; (2) the code is written
in such a way that additional terms can be added easily. The coding
corresponds one-by-one to the analytic formulas in most places except
where efficiency demands reordering the calculations.

\subsubsection{The single-particle basis}

In a mean field theory one seeks to describe the many-body system
exclusively in terms of a set of single-particle wave functions
$\psi_\alpha$ with fractional occupation amplitudes $v_\alpha$, i.e.
\begin{subequations}
  \begin{equation}
    \left\{\psi_\alpha,v_\alpha,\alpha=1,...,\Omega\right\}
  \end{equation}
  where $\Omega$ denotes the size of the active s.p. space.  The
  occupation amplitude can take values continuously in the interval
  $[0,1]$. The complementary non-occupation amplitude is
  $u_\alpha=\sqrt{1-v_\alpha^2}$. A formal definition of the BCS
  mean-field state reads
  \begin{equation}
    |\Phi\rangle
    =
    \prod_{\alpha>0}\big(
    u_\alpha^{\mbox{}}+v_\alpha^{\mbox{}}\hat{a}^+_\alpha\hat{a}^{+}_{\bar\alpha}
    \big)|0\rangle
    \label{eq:BCState}
  \end{equation}
\end{subequations}
where $|0\rangle$ is the vacuum state, $\hat{a}^+_\alpha$ the
generator of a Fermion in state $\psi_\alpha$, and $\bar\alpha$ the
time reverse partner to state $\alpha$.  We will use variation of the
BCS amplitudes $v_\alpha^{\mbox{}}$ only in the static part of
even-even nuclei where the time reverse partner is unambiguously
defined. These amplitudes are kept frozen during
dynamical propagation whenever BCS pairing is used for a nucleus.

\subsubsection{Local densities and currents \label{sec:dens}}

The Skyrme-energy-density functional is defined in terms of only a
few local densities and currents. These are the {\em time-even} fields
\begin{align}
   \rho_q&=\displaystyle
    \sum_{\alpha\in q}\sum_{s}  
    v_{\alpha}^2|\psi_{\alpha}(\vec{r},s)|^2
    &&
    \text{density}
   \notag \\
   \vec{J}_q &=\displaystyle
    -\I\sum_{\alpha\in q}\sum_{ss'} v_{\alpha}^2
    \psi_{\alpha}^*(\vec{r},s)
    \nabla\! \times\! \vec{\sigma}_{ss'} 
    \psi^{\mbox{}}_{\alpha}(\vec{r},s')
    &&
    \text{spin-orbit dens.}
    \notag\\
    \tau_q&=\displaystyle
    \sum_{\alpha\in q}\sum_{s}  
    v_{\alpha}^2|\nabla\!\psi_{\alpha}(\vec{r},s)|^2
    &&
    \mbox{kinetic density,}
 \label{eq:rtjeven}
\end{align}
the {\em time-odd} fields
\begin{align}
 \vec{s}_q&=\displaystyle
    \sum_{\alpha\in q}\sum_{ss'} v_{\alpha}^2
    \psi^*_{\alpha}(\vec{r},s)\vec{\sigma}_{ss'}
    \psi^{\mbox{}}_{\alpha}(\vec{r},s')
    &&
    \mbox{spin density}
   \notag\\
   \vec{\jmath}_q&=\displaystyle
    \Im{m}\left\{\sum_{\alpha\in q}\sum_{s}    v_{\alpha}^2 
      \psi^*_{\alpha}(\vec{r},s)
      \nabla\!\psi^{\text{}}_{\alpha}(\vec{r},s)\right\}
    &&
    \text{current density,}
 \label{eq:rtjodd}
\end{align}
and a field with undefined time parity:
  \begin{align}
    \xi_q 
    &=\displaystyle
    {\sum_{\alpha\in q}\sum_{s}    u_{\alpha}v_{\alpha} 
      \psi_{\overline{\alpha}}(\vec{r},s)
      \psi_{\alpha}}(\vec{r},s)
    &&
    \text{pairing density}
 \label{eq:rtjpair}
\end{align}
where $q$ labels the nucleon species with $q=p$ for protons and $q=n$
for neutrons. A local density/current without $q$ index stands for the
total quantity, e.g.  $\rho=\rho_p+\rho_n$ is the total density, and
similarly for the other densities/currents. 

\subsubsection{The energy-density functional}\label{sec:efunc}

The mean-field equations solved in the code are based on the widely
used Skyrme energy functional. For recent reviews see
\cite{Ben03aR,Erl11a}.  The functional at the level at which it is
used here can be written as
\begin{subequations}
  \label{eq:efunc}
  \begin{equation}
    \label{eq:efundet}
    E_\mathrm{tot}
    =
    T+(E_0+E_1+E_2+E_3+E_\mathrm{ls})+E_{\rm Coulomb}+E_{\rm pair}+E_{\rm corr},
  \end{equation}
  where the parentheses were used to group the terms arising from the
  Skyrme force. The various terms read in detail
  \begin{itemize}
  \item $T$: the total kinetic energy calculated as
    \begin{equation}
      \label{eq:ehft}
      T=\sum_q \frac{\hbar^2}{2m_q}\int\D^3 r\, \tau_q(\vec r)
    \end{equation}
    with $\tau_q$ the isospin-specific kinetic density of Eq.~(\ref{eq:rtjeven}).
  \item $E_0$: The $b_0$ and $b'_0$-dependent part is
    \begin{equation}
      \label{eq:ehf0}
      E_0=\int \D^3\!r\,\left(\frac{b_0}{2}\rho^2-\frac{b_0'}{2}\sum_q\rho_q^2\right).
    \end{equation}
  \item $E_{1}$: kinetic terms containing the coefficients $b_1$ and $b_1'$:
   \begin{equation}
      \label{eq:ehf1}
      E_1=\int \D^3\!r\,\left(b_1[\rho\tau-{\vec\jmath\,}{}^2]
        -b_1'\sum_q[\rho_q\tau_q-{\vec\jmath_q}{}^2]\right).
    \end{equation}
  \item $E_{2}$: terms containing the coefficients $b_2$ and $b_2'$.
    They involve the Laplacians of the densities.
    \begin{equation}
      \label{eq:ehf2}
      E_2=\int \D^3\!r\,\left(-\frac{b_2}{2}\rho\Delta\rho
        +\frac{b_2'}{2}\sum_q\rho_q\Delta\rho_q\right).
    \end{equation}
  \item $E_{3}$: The many-body contribution is given by
    \begin{equation}
      \label{eq:ehf3}
      E_{3}=\int \D^3\!r\,\left(\frac{b_3}{3}\rho^{\alpha+2}
        -\frac{b_3'}{3}\rho^\alpha\sum_q\rho_q^2\right).
    \end{equation}
  \item $E_\mathrm{ls}$: the spin-orbit energy
    \begin{equation}\begin{split}
      E_\mathrm{ls}&=\int \D^3\!r\,\biggl(
        -b_4[\rho\nabla\cdot\vec J+
        \vec s\cdot(\nabla\times\vec\jmath)]\\
        &-b_4'\sum_q[\rho_q\nabla\cdot\vec J_q+
        \vec s_q\cdot(\nabla\times\vec\jmath_q)]\biggr)
      \label{eq:ehfls}
    \end{split}
  \end{equation}
 \item $E_\mathrm{C}$: the Coulomb energy. It consists of the
    standard expression for a charge distribution in its own field
    (Hartree term) plus the exchange term in the Slater approximation
   ~\cite{Sla51}. The formula is
    \begin{equation}
      \label{eq:ehfc}
      E_\mathrm{C}
      =
      \frac{e^2}{2}\int \D^3\!r \D^3\!r'
      \frac{\rho_p(\vec{r})\rho_p(\vec{r'})}{|\vec{r}-\vec{r'}|}
      -
      \int \D^3\!r\,\frac{3e^2}{4}\left(\frac3{\pi}\right)^{\tfrac1{3}}\rho_p^{4/3}
    \end{equation}
    where $e$ is the elementary charge with $e^2=1.43989$ MeV\,fm.
  \item $E_\mathrm{pair}$: the pairing energy. It consists of a
    contact pairing interaction involving the pairing densities $\xi_q$
    augmented by an optional density dependence. The formula is
    \begin{equation}
      E_{\rm pair}
      =
      \frac{1}{4} \sum_{q\in\{p,n\}}V_\mathrm{pair,q}
      \int \D^3r \xi^2_q
      \left[1 -\frac{\rho}{\rho_{0,\mathrm{pair}}}\right]\;.
      \label{eq:epair}
    \end{equation}     
    It contains a continuous switch, the parameter
    $\rho_{0,\mathrm{pair}}$. A pure $\delta$-interaction (DI), also
    called volume pairing, is recovered for
    $\rho_{0,\mathrm{pair}}\longrightarrow\infty$. The general case is
    the density dependent $\delta$-interaction (DDDI).  A typical
    value near matter equilibrium density
    $\rho_{0,\mathrm{pair}}=0.16$ fm$^{-3}$ concentrates pairing to
    the surface. The most flexible choice is to consider
    $\rho_{0,\mathrm{pair}}$ as an additional free parameter. Actual
    adjustments with this option deliver a form of the pairing
    functional which stays in between the extremes of volume and
    surface pairing~\cite{Klu09a}. 
  \end{itemize}
\end{subequations}

The term $E_{\rm corr}$ stands for all additional corrections from
correlations beyond the mean field that might be added.  Most calculations
include at least the center-of-mass correction $E_\mathrm{cm}$.
For deformed nuclei this should be augmented by a rotational correction and
for soft nuclei by correlations from all low-energy quadrupole motions
\cite{Klu08a}. So far, these correlations are usually added a
posteriori after static calculations. A fully variational treatment
and a dynamical propagation of the c.m. correction is extremely
involved and usually not considered. We aim predominantly at dynamical
simulations and thus here ignore all such correlations, using
\begin{equation}
  E_\mathrm{cm}=0
  \quad.
\end{equation}
This has to be taken into account when comparing ground state energies
from this code with results from other static codes which contain a
c.m.\ correction $E_\mathrm{cm}\neq 0$. However, one has to take into
account that there are two different ways to perform the c.m.\ 
correction. They are selected by the parameter {\tt zpe}. 
One strategy is to augment the energy by
$E_\mathrm{cm}$ a posteriori. We associate this option with
{\tt{zpe}}=1.
This means that in our code nothing is done. The other strategy is to
modify the nucleon mass by $m\longrightarrow m-m/A$ and to not have a
posteriori correction. This way is chosen in a couple of traditional
parametrizations, e.g., in SkM$^*$~\cite{Bar82a}. We keep this option
for consistency and associate it with {\tt zpe}=1. But note that this
choice runs into inconsistencies in collisions and fragmentation as
it employs only the total mass number $A$ and cannot account for the
masses of the fragments.

The functional in the above form contains the minimal number of terms
which are needed to guarantee Galilean invariance~\cite{Eng75a,Erl11a}
and so to allow performance of TDHF calculations which respect all basic
conservation laws. We ignore the tensor spin-orbit terms and spin-spin
couplings~\cite{Ben03aR,Per04a,Erl11a}. These may be important for
magnetic excitations~\cite{Erl11a} and odd nuclei~\cite{Pot10a} which
are, however, not the prime focus of TDHF studies.

\subsubsection{Force coefficients}
The above formulation in terms of the Skyrme energy functional
introduces the force parameters $b_0$, $b'_0$, ...  $b'_4$ naturally
as the factors in front of each contribution in the terms
(\ref{eq:ehf0}). Traditionally, the functional is deduced from a
Skyrme force which is a density-dependent, zero-range interaction
\cite{Vau72a}.  The $t$ and $x$ coefficients in this Skyrme-force
definition are related to the $b$ coefficients in the functional
definition as
\begin{equation}
  \label{eq:force-coeff}
  \begin{split}
    b_0&=t_0\,\left(1+\tfrac1{2}x_0\right) \\
    b_0'&=t_0\,\left(\tfrac1{2}+x_0\right)\\
    b_1&=\tfrac1{4}\left[t_1\,\left(1+\tfrac1{2}x_1\right)+
      t_2\,\left(1+\tfrac1{2}x_2\right)\right] \\
    b_1'&=\tfrac1{4}\left[t_1\,\left(\tfrac1{2}+x_1\right)-
      t_2\,\left(\tfrac1{2}+x_2\right)\right] \\
    b_2&=\tfrac1{8}\left[3t_1\,\left(1+\tfrac1{2}x_1\right)-
      t_2\,\left(1+\tfrac1{2}x_2\right)\right] \\
    b_2'&=\tfrac1{8}\left[3t_1\,\left(\tfrac1{2}+x_1\right)+
      t_2\,\left(\tfrac1{2}+x_2\right)\right]\\
    b_3&=\tfrac1{4}t_3 \left(1+\tfrac1{2}x_3\right) \\
    b_3'&=\tfrac1{4}t_3 \left(\tfrac1{2}+x_3\right) \\
    b_4&=\tfrac1{2}t_4
  \end{split}
\end{equation}
The coefficient $b_4'$ is usually fixed as $b'_4=\frac{1}{2}t_4$ for most
traditional Skyrme forces. More recent variants of Skyrme forces (SkI3
etc.) handle it as a separate free parameter~\cite{Rei95a}. In addition
to the $b$ and $b'$ parameters, there is the power coefficient in the
(originally) three-body term, which is usually called $\alpha$, but in
the code is referred to as {\tt power}.  The input of the force to the
code is done in terms of the $t_i$, $x_i$ parameters.

\subsubsection{The single-particle Hamiltonian}
\label{sec:otherfields}

The mean-field Hamiltonian $\hat{h}$ is derived from the energy
functional of Section~\ref{sec:efunc} by variation
$\partial_{\psi_\alpha^*}E=\hat{h}\psi_\alpha$.  It reads in detail
\begin{subequations}
  \label{eq:mfpots}
  \begin{equation}\begin{split}
    \label{eq:spham}\nonumber
    \hat h_q& =
    U_q(\vec r)-\nabla\cdot\left[B_q(\vec r)\nabla\right]
    +\I\vec W_q\cdot(\vec\sigma\times\nabla)
    +\vec S_q\cdot\vec\sigma\\
    &-\frac{\I}{2} \left[(\nabla\cdot\vec A_q)+2\vec
      A_q\cdot\nabla\right]\,,
  \end{split}
\end{equation}
 with $q\in\{p,n\}$ as usual distinguishing the different
  Hamiltonians for protons and neutrons.  For the protons the Coulomb
  potential is also added.  The first term is the local part of the
  mean field, which acts on the wave functions like a local
  potential. It is defined as
  \begin{equation}\begin{split}
    U_q
    &=
    b_0\rho-b_0'\rho_q+b_1\tau-b_1'\tau_q-b_2\Delta\rho
    +b_2'\Delta\rho_q
    \nonumber\\
    &+b_3\tfrac{\alpha+2}{3}\,\rho^{\alpha+1}-b_3'\tfrac2{3}\,\rho^\alpha\rho_q
    -b_3'\tfrac\alpha{3}\,\rho^{\alpha-1}\sum_{q'}\rho_{q'}^2
    \nonumber\\
    &
    -b_4\nabla\cdot\vec J-b_4'\nabla\cdot \vec{J}_q
    \;. 
    \label{eq:upot}
  \end{split}
\end{equation}
Next comes the ``effective mass'', which replaces the standard
  $\tfrac{\hbar^2}{2m}$ factor by the isospin and space-dependent
  expression
  \begin{equation}
    \label{eq:bmass}
    B_q=\frac{\hbar^2}{2m_q}+b_1\rho-b_1'\rho_q.
  \end{equation}
  Note that the Skyrme force definitions contain the first term
  (nucleon mass) as a parameter which varies slightly from
  parametrization to parametrization and may be different for protons
  and neutrons. The spin-orbit potential is
  \begin{equation}
    \vec W_q
    =
    b_4\nabla\rho+b_4'\nabla\rho_q
    \quad.
    \label{eq:wlspot}
  \end{equation}
  The above three terms involve the time-even densities. Dynamical
  effects come into play with the next terms which include the
  time-odd contributions from current and spin-density:
  \begin{align}
    \label{eq:aq}
    \vec A_q&=-2b_1\vec\jmath+2b_1'\vec\jmath_q-b_4\nabla\times\vec s
    -b_4'\nabla\times\vec s_q.
    \;, \\
   \label{eq:spot}
    \vec S_q&=-b_4\nabla\times\vec\jmath-b_4'\nabla\times\vec\jmath_q
    \;.
\end{align}
\end{subequations}

\subsection{Coupling to external fields}
\label{sec:external}

For the dynamic case, the system can also be coupled to an external
excitation field, to study collective response such as in giant
resonances. The present code only implements a very simple case, since
it is expected that most serious applications will need modifications,
which are quite easy to incorporate.

The external field is introduced as a time-dependent, local operator
\begin{subequations}
  \begin{equation}
    \hat{h}_q\longrightarrow\hat{h}_q+U_{q,\mathrm{ext}}(\vec{r},t)
    \quad,\quad
    U_{q,\mathrm{ext}}
    =
    \eta\,f(t)\,F_q(\vec r)\;,
  \end{equation}
  where $f(t)$ carries the temporal profile of the excitation
  mechanism, $F_q(\vec r)$ is some local operator, and $\eta$ tunes
  the overall strength.  The spatial distribution $F_q(\vec{r})$, is
  allowed to be different for the two isospins $q$. Typical examples
  are isoscalar and isovector multipole operators as, e.g., the
  isoscalar quadrupole $F_q(\vec{r})=2z^2-x^2-y^2$.

  The prefactor $\eta$ is a strength parameter which allows
  scanning different excitation strengths easily while keeping the
  temporal and spatial profiles the same. It should be noted that the
  absolute magnitude of the perturbing potential by itself usually has
  little direct meaning. What counts is the excitation energy caused
  by the perturbation and subsequently the magnitude of vibrations in
  the observables (such as the time-dependent quadrupole moment). An
  exception is, e.~g., the simulation of the close approach of another
  nucleus that stays external to the computational grid, where the
  potential is uniquely defined.

  One important point remains to be noted concerning the spatial
  profile $F_q(\vec{r})$. This can be illustrated by the quadrupole
  operator. Let us assume an instant where $A>0$ and $f(t)>0$.  For
  then the operator $\propto 2z^2-x^2-y^2$ leads to a perturbing potential
  $U_\mathrm{ext}$ which is binding in $z$-direction but
  asymptotically unbound in the $x$- and $y$-directions. This can cause
  unphysical effects in case of large strengths and/or numerical boxes.
  For this reason it is useful to have a cut-off by a
  Woods-Saxon like function according to~\cite{Rut95a}
  \begin{equation}
    \label{eq:extdamp} §§§
    F_q(\vec r)
    \rightarrow 
    \frac{F_q(\vec r)}{1+e^{(r-r_0)/\Delta r}}\;,
  \end{equation}
  where $r_0$ and $\Delta r$ are parameters describing a transition
  region sufficiently outside the nucleus, but also sufficiently small
  to maintain binding.

  Another problem associated with the external field is that in
  general it will not be periodic but instead have discontinuities on
  the boundary when crossing into the neighboring cell. If damping is
  sufficiently strong, the field may be practically zero on the
  boundary and thus become periodic. Another solution for this problem
  is to make the field explicitly periodic by replacing the
  coordinates with periodic substitutes. The exact formulation depends
  on the specific field used; for the above-mentioned quadrupole
  operator, which depends only on the squares of the coordinates,
  e.~g., substituting
  \begin{equation}\label{eq:extperiodic}
    x^2\rightarrow\sin^2\left(\pi x/x_L\right),
  \end{equation}
  with $x_L={\tt nx}\,\Delta x$ the period interval, will provide the
  proper behavior as the sine squared has a period of $\pi$ and there
  is no unphysical decrease of this function near the boundary. Of
  course the analogous transformation has to be applied to $y$ and
  $z$.

  The time dependence $f(t)$ is modeled as a short pulse centered
  around some excitation frequency $\omega$. The code allows a choice
  between two pulse envelopes:
  \begin{enumerate}
  \item A Gaussian of the form
    \begin{equation}
      \label{eq:extgauss}
      f(t)=\exp\left(-(t-\tau_0)^2/\Delta\tau^2\right)\cos(\omega(\tau-\tau_0))\;,
    \end{equation}
    with peak time $\tau_0$ and width $\Delta\tau$.
  \item A cosine squared function defined via
    \begin{equation}
      \label{eq:extcos}
      f(t)=\cos\left(\frac\pi{2}\left(\frac{t-\tau_0}{\Delta\tau}\right)^2\right)
      \theta\left(\Delta\tau-|t-\tau_0|\right)
      \cos(\omega(\tau-\tau_0))\;,
    \end{equation}
    which is confined to the intervals $t\in
    (\tau_0-\Delta\tau,\tau_0+\Delta\tau)$. This envelope is also
    characterized by a peak time of $\tau_0$ and width $\Delta\tau$.
  \end{enumerate}
\end{subequations}
Broad envelopes provide a high frequency resolution and so concentrate
the excitation around the driving frequency $\omega$. Short pulses
lose frequency selectivity and excite a broad band of frequencies.

The extreme case is an infinitely short pulse,
$\Delta\tau\longrightarrow{0}$. It amounts eventually to an
instantaneous boost of the initial wave functions which can be
expressed as a phase factor according to
\begin{equation}
  \label{eq:extboost}
  \psi_k(\vec r,s,t\!=\!0)
  =
  \psi_{k,0}(\vec r,s)\,\exp\left(-\I \eta F_q(\vec r)\right)\;,
\end{equation}
where $\psi_{k,0}$ stands for the stationary wave function before
boost. This instantaneous boost, being infinitely short, is
insensitive to the problem of asymptotically unstable potentials and
allows the use of a driving field $F_q$ without damping
  (\ref{eq:extdamp})
which, in turn, simplifies spectral analysis (see Section
\ref{sec:collex}).

The effect of the boost (\ref{eq:extboost}) can be understood by
virtue of the Madelung decomposition of a complex wave function
$\phi(\vec r)=\chi(\vec r)\exp(\I S(\vec r))$ with $\chi$ and $S$
being real. A straightforward calculation leads to the probability
flow density as $\vec{\jmath}=\hbar\chi^2\nabla S/m$. Dividing by the
density $\chi^2$ then produces the ``probability-flow velocity''
$\vec{v}=\hbar\nabla S/m$. An illustrative example is the plane wave
where $\chi=$constant and $S=\vec{k}\cdot\vec{r}$ which yields the
correct result $\vec{v}=\hbar\vec{k}/m$.  In the present case,
assuming that the static wave functions themselves can be written as
real functions, we get an initial velocity $\vec{v}=-\nabla F_q$,
i.~e., just in the direction of the classical force resulting from the
``velocity potential'' $F_q(\mathbf{r})$.

\subsection{Static Hartree-Fock}
\label{sec:static}

\subsubsection{The coupled mean-field and BCS equations}

The stationary equations are obtained variationally. Variation with
respect to the single-particle wave functions $\psi_{\alpha}$ yields
the mean field equations~\cite{Gre05aB,Mar10aB}
\begin{subequations}
  \label{eq:mfeq}
  \begin{equation}
    \hat{h}\psi_{\alpha}=\varepsilon_\alpha\psi_{\alpha}\;,
    \label{eq:statmfeq}
  \end{equation}
  where $\hat{h}$ is the mean-field Hamiltonian (\ref{eq:spham}) and
  $\varepsilon_\alpha$ is the single-particle energy of state
  $\alpha$.  Simultaneous variation of $\psi_{\alpha}$ together with
  the occupation amplitude $v_\alpha$ yields the
  Hartree-Fock-Bogolyubov equations~\cite{Gre05aB,Rin80aB,Ben03aR}
  which complicate Eq.~(\ref{eq:statmfeq}) by non-diagonal terms on
  the right-hand-side~\cite{Rei97a}. We employ here the BCS
  approximation which exploits time-reversal symmetry where each
  single-particle state has a time reversed partner
  $\psi_{\alpha}\leftrightarrow\psi_{\overline{\alpha}}$ as was
  already implied in the pairing densities $\xi_q$ in Eq.~(\ref{eq:rtjpair}).
  Each pair of time-conjugate states is associated with an occupation
  amplitude $v_\alpha$. These are determined by the BCS equation
  $\left(\varepsilon_\alpha-\epsilon_{\mathrm{F},q_\alpha}\right)
  \left(u_\alpha^2-v_\alpha^2\right) = \Delta_\alpha
  u_\alpha^{\mbox{}}v_\alpha^{\mbox{}} $ which can be resolved to a
  closed expression for the occupation amplitudes as
  \begin{eqnarray}
    v_\alpha^2
    & =&
    \frac{1}{2} \left( 1 -
      \frac{\varepsilon_\alpha-\epsilon_{\mathrm{F},q_\alpha}}
      {\sqrt{(\varepsilon_\alpha-\epsilon_{\mathrm{F},q_\alpha})^2
          +\Delta_\alpha^2}}
    \right)
    \quad,
    \label{eq:va}\\
    \Delta_\alpha 
    &
    =
    &
    \frac{1}{2}V_\mathrm{pair,q_\alpha}
    \int \D^3r\,\psi^+_\alpha\psi_\alpha^{\mbox{}}
    \xi^{\mbox{}}_{q_\alpha}
    \left[1 -\frac{\rho}{\rho_{0,\mathrm{pair}}}\right]
    \quad,
    \\
    \epsilon_{\mathrm{F},q}
    &:&
    \sum_{\alpha\in q}v_\alpha^2=N_q
    \quad.
    \label{eq:parnum}
  \end{eqnarray}
  $q_\alpha$ denotes the nucleon type to which state $\alpha$
  belongs, ${\alpha\in q}$ means all states of type $q$, and $N_q$ is
  the number of nucleons of type $q$ (identified as $N_p=Z$ and
  $N_n=N$).  The Fermi energies $\epsilon_{\mathrm{F},q_\alpha}$ serve
  to regulate the average particle number to the required values
  $N_q$.
\end{subequations}
Here, the space of pairing-active states is just the space of states
actually included in the calculation. The results of BCS pairing
depend slightly on the size of the active space
\cite{Gre05aB,Rin80aB}.  We recommend using about 
$$N_q+\frac{5}{3}N_q^{2/3}$$
single-nucleon states, which comes closest to the dynamical pairing
space of Ref.~\cite{Ben00a}.

\subsubsection{Iterative solution}
\label{sec:statiter}

The coupled mean-field and BCS equations (\ref{eq:mfeq}) are solved
iteratively. The wave functions are iterated with a gradient step which
is accelerated by kinetic-energy damping~\cite{Rei82a,Blu92a}
\begin{equation}
  \psi_\alpha^{(n+1)}
  =
  \mathcal{O}\left\{
    \psi_\alpha^{(n)} 
    - 
    \frac{\delta}{\hat{T} + E_0} 
    \left( \hat{h}^{(n)} - 
      \langle\psi_\alpha^{(n)}|\hat{h}^{(n)}|\psi_\alpha^{(n)}\rangle
    \right)\psi_\alpha^{(n)}\right\}
  \label{eq:dampstep}
\end{equation}
where $\hat{T}=\hat{p}^2/(2m)$ is the operator of kinetic energy,
$\mathcal{O}$ means orthonormalization of the whole set of new wave
functions, and the upper index indicates the iteration number.  Note
that this sort of kinetic-energy damping is particularly suited for
the fast Fourier techniques that we use in the present code.  The
damped gradient step has two numerical parameters, the step size
$\delta$ and the damping regulator $E_0$. The latter should be chosen
typically of the order of the depth of the local potential $U_q$.  In
practice, we find $E_0=100$ MeV a safe choice. The step size is of
order of $\delta=0.1...0.8$. Larger values yield faster iteration, but
can run more easily into pathological conditions. The optimum
choice depends somewhat on the Skyrme parametrization. Those with
effective mass $m^*/m\approx 1$  allow larger $\delta$ values. Low
$m^*/m$ may require a reduction in the step size.

After each such wave function step, the BCS equations 
(\ref{eq:va}--\ref{eq:parnum}) are solved with
$\varepsilon_\alpha=\langle\psi_\alpha|\hat{h}|\psi_\alpha\rangle$,
the densities are updated (Eqs.~\ref{eq:rtjeven}-\ref{eq:rtjpair}), 
and new mean fields computed.
(\ref{eq:mfpots}). This then provides the starting point for the next
iteration. The process is continued until sufficient convergence is
achieved. We consider as the convergence criterion the average energy
variance, or fluctuation, of the single particle
states
\begin{subequations}
  \label{eq:totvar}
  \begin{eqnarray}
    \overline{\Delta\varepsilon}
    &=&
    \sqrt{\frac{\sum_\alpha\Delta\varepsilon_\alpha^2}{\sum_\alpha 1}}
    \quad,
    \\
    \Delta\varepsilon_\alpha^2
    &=&
    \langle\psi_\alpha|\hat{h}^2|\psi_\alpha\rangle
    -
    \varepsilon_\alpha^2
    \quad,
    \label{eq:spvar}
    \\
    \varepsilon_\alpha
    &=&
    \langle\psi_\alpha|\hat{h}|\psi_\alpha\rangle
    \quad.
    \label{eq:spenerg}
  \end{eqnarray}
\end{subequations}
The single particle energy $\varepsilon_\alpha$ is defined here as an
expectation value. It finally becomes an eigenvalue in 
Eq.~(\ref{eq:statmfeq}) if the iteration has converged to
$\Delta\varepsilon_\alpha\approx 0$.  Vanishing total variance
$\overline{\Delta\varepsilon}$ signals that we have reached minimum
energy, i.e. a solution of the mean-field plus BCS equations.  One has
to be aware, however, that this may be only a local minimum (isomeric
state). It requires experience to judge whether one has found the
absolute energy minimum. In case of doubt, one should redo a couple of
static iterations from very different initial configurations.

This raises the question of how to initialize the iteration.  We take
as a starting point the wave functions of the deformed harmonic
oscillator (see point \ref{it:iniho} in Section~\ref{sec:init}). These
are characterized by $\vec{n}=(n_x,n_y,n_z)$, the number of nodes in
each direction. We stack the wave functions in order of increasing
oscillator energy
$\epsilon_\alpha^{(0)}=\hbar\omega_xn_x+\hbar\omega_yn_y+\hbar\omega_zn_z$
and stop if the desired number of states is reached. The deformation of
the initializing oscillator influences the initial state in two ways:
first, through the deformation of the basis wave functions as such,
and second, through the energy ordering of the $\epsilon_\alpha^{(0)}$
and corresponding sequence of levels built. Variation of initial
conditions means basically a variation of the oscillator deformation.
For example, the iteration will most probably end up in a prolate
minimum if the initial state was sufficiently prolate, and in an
oblate minimum after an oblate initial state. It depends on the
nucleus which one is the absolute minimum.

\subsection{TDHF}

\subsubsection{The time-dependent mean-field equations}

The TDHF equations are determined from the variation of the action
\begin{equation*}
  S=\int \D t\left[E[\{\psi_\alpha\}]-
    \sum_\alpha\langle\psi_\alpha|\I \partial_t|\psi_\alpha\rangle
  \right]\;,
\end{equation*}
with respect to the wave functions $\psi_\alpha^+$ where the energy is
given as in Eq.~(\ref{eq:efunc})~\cite{Mar10aB}.  This yields the TDHF
equation
\begin{equation}
  \I \partial_t\psi_{\alpha}=\hat{h}\psi_{\alpha}\;,
  \label{eq:dynmfeq}
\end{equation}
where $\hat{h}$ is, again, the mean-field Hamiltonian
(\ref{eq:spham}).  For simplicity, we are not considering variation of
the occupation amplitude in the time-dependent case. The occupation
amplitudes obtained from static iteration are kept frozen during time
evolution. For studies of mean-field flow at moderate
excitations (heavy-ion collisions, giant resonances) this
approximation is legitimate. However, a study of truly low energy
dynamics in the range of a few MeV (soft vibrations, fission) requires
a full time-dependent Hartree-Fock-Bogolyubov treatment and should not
be undertaken with the present code.

\subsubsection{Time development algorithm}
\label{sec:timevol}

The TDHF equation (\ref{eq:dynmfeq}) can be formally resolved into an
integral equation as
\begin{subequations}
  \begin{eqnarray}
    |\psi_\alpha(t\!+\!\Delta t)\rangle
    &=&
    \hat U(t,t\!+\!\Delta t)|\psi_\alpha(t)\rangle
    \\
    \hat U(t,t\!+\!\Delta t)
    &=&
    \hat{\mathcal{T}}\exp\left(-\frac{\I}{\hbar}\int_t^{t+\Delta t}
      \hat h(t')\,\D t'\right)
    \;,
    \label{eq:timexp}
  \end{eqnarray}
\end{subequations}
where $\hat U$ is the time-evolution operator and $\hat{\mathcal{T}}$
the time-ordering operation.  This time evolution is unitary, thus
conserving orthonormalization of the single-particle wave functions,
and it conserves the total energy (\ref{eq:efunc}) provided that there are
no time-dependent external fields.  To convert this involved
operator into an efficiently computable but also sufficiently accurate
form a predictor-corrector strategy is used:
\begin{enumerate}
\item In a first step (predictor), we determine the single-particle
  Hamiltonian at midtime $\hat{h}(t\!+\!\Delta t/2)$. 
  To that end, a trial step by $\Delta t$
  \begin{equation}
    \tilde{\psi}_\alpha
    =
    \exp\left(-\tfrac{\I}{\hbar}
      \hat h(t)\,\Delta t\right)\psi_\alpha(t)
    \label{eq:midstep}
  \end{equation} 
  is performed using the mean field $\hat{h}(t)$ at initial time $t$.
  This produces an estimate $\tilde{h}$ of the mean field Hamiltonian at $t+\delta
  t$. The Hamiltonian ad midtime is then obtained as the average
  $\hat{h}(t+\delta{t})=\left(\hat{h}(t)+\tilde{h}\right)/2$.
 
\item In a second step (corrector),  the trial-step wave
  functions $\{\tilde{\psi}_\alpha\}$ are used to compute an estimate
  (predictor) of the mean-field at the final time $\hat{h}_\mathrm{pre}$
  in the standard manner by accumulating the densities and composing
  these to the Skyrme mean-field Hamiltonian (\ref{eq:mfpots}). The
  predictor for the Hamiltonian at midtime $t\!+\!\Delta t/2$ then
  becomes
  $\tilde{h}\approx\left(\hat{h}(t)+\hat{h}_\mathrm{pre}\right)/2$.
  This is used to finally perform a full time step (again with frozen
  Hamiltonian, but now $\tilde{h}$)
  \begin{equation}
    \psi_\alpha(t\!+\!\Delta t)
    =
    \exp\left(-\tfrac{\I}{\hbar}\tilde{h}\,\Delta t\right)
    \psi_\alpha(t)
    \;.    
    \label{eq:finstep}
  \end{equation}
\item In both cases the operator exponential is evaluated by a Taylor
  series expansion up to order $m$:
  \begin{equation}
    \exp\left(-\tfrac{\I}{\hbar}\hat{h}\,\Delta t\right)\psi
    \approx
    \sum_{n=0}^m
    \frac{(-\I\Delta t)^n}{\hbar^n\,n!}\hat h^n\psi\;,
    \label{eq:timepower}
  \end{equation}
  where $\hat{h}$ is the actual mean field in step (\ref{eq:midstep}),
  or (\ref{eq:finstep}) respectively.  $\hat h^n\psi$ is computed
  in straightforward manner by successive application of the mean
  field Hamiltonian, i.e.  $\hat
  h^n\psi=\underbrace{\hat{h}(...(\hat{h}}_{n \mbox{\tiny
      times}}\psi)...)$.
\end{enumerate}
The Taylor expansion spoils strict unitarity of the exponential
$\exp\left(-\tfrac{\I}{\hbar}\hat{h}\,\Delta t\right)$ and energy
conservation. We turn this flaw into an advantage and use norm
conservation as well as energy conservation (if it applies) as
counter-check of the quality of the step along the propagation. The
reliability depends, of course, on a proper choice of the numerical
parameters in this step which are the step size $\Delta t$ and the
order of the Taylor expansion $m$. The step size is limited by the
maximum possible kinetic energy and by the typical time scales of changes in
the mean field $\hat{h}$. The maximum kinetic energy, in turn, depends
on the grid spacing as $\propto \Delta x^{-2}$. A choice of $\Delta
t=0.1-0.2$fm/c is applicable in connection with $\Delta x=0.7-1$/fm.
For the order of Taylor expansion, one needs at least $m=4$. One may
also consider $m=6$. Higher orders $m$ are not necessary for
the typical values of $\Delta t$.

\subsubsection{Collective excitations}
\label{sec:collex}

Giant resonances are prominent excitation modes of nuclei.  Best known
is probably the isovector giant dipole resonance,  but there are many more
modes depending on isospin
and angular momentum. The typical resonance energies lie in a region
from 10 to 30 MeV where the present TDHF code with frozen occupation
numbers is applicable because the relevant energy range lies far above
the pairing gap (1--2 MeV).  The generation of these modes is
particularly simple within the present TDHF treatment. One first
produces a stationary state as outlined in Section~\ref{sec:static}
and then triggers the excitation by a time-dependent external
field as described in Section~\ref{sec:external}. A broad pulse allows
triggering particular excitation energies. An infinitely short pulse
amounts to an instantaneous boost.

The boost is a generic excitation of a system which gives the same
weight to all frequencies. It is thus ideally suited for analyzing in
an unbiased manner the excitation spectra of a system.  This, in turn,
allows a thorough spectral analysis.  To obtain the spectral
distribution of isovector dipole strength, one applies a boost with
small strength $\eta$ and $F_q=\hat{D}\propto r^1Y_{10}\tau_z$ the
isovector dipole operator. The Slater determinant
$|\Phi(t)\rangle$ is propagated in TDHF for a sufficiently long time while recording
the dipole moment $D(t)=\langle\Phi(t)|\hat{D}|\Phi(t)\rangle$. The
dipole strength is finally extracted from the Fourier transform
$\tilde{D}(\omega)$ as
$S_D(\omega)=\Im\left\{\tilde{D}(\omega)\right\}/\eta$.  The
straightforward Fourier transform leads to artifacts if the dipole
signal has not fully died out at the end of the simulation time.  In
the general case, some filtering is necessary to suppress artifacts
from cutting the signal at a certain final time~\cite{Pre92aB}. In
practice, it is most convenient to use filtering in the time domain by
damping the signal $D(t)$ towards the final time.  A robust choice is
\begin{equation}
  D(t)
  \quad\longrightarrow\quad
  D_\mathrm{fil}(t)
  =
  D(t)\cos\left(\frac{\pi}{2}\frac{t}{t_\mathrm{final}}\right)^{n_\mathrm{fil}}
\end{equation}
where $t_\mathrm{final}$ is the final time of the simulation. This
guarantees that the effective signal $D_\mathrm{fil}$ vanishes at the
end of the interval. The $\cos^n$ profile switches off gently and
leaves as much as possible from the relevant signal at early times.
The parameter $n_\mathrm{fil}$ determines the strengths of filtering.
Value of order of 4--6 are recommended to suppress the
artifacts safely.  For a detailed description of this spectral analysis see
\cite{Cal97a}. For typical applications in nuclear physics see
\cite{Rei07a}. It is to be noted that the code does not include this
final step of spectral analysis. The time dependent signals are
printed on the protocol files {\tt monopoles.res}, {\tt dipoles.res},
and {\tt quadrupoles.res}. It is
left to the user to perform the final steps towards a spectral
distribution.  A word is in order about $t_\mathrm{final}$. It
determines the resolution of the spectral analysis. The corresponding
energy bins are given by
$\delta{E}_\mathrm{exc}=\hbar\pi/t_\mathrm{final}$. Windowing effectively
reduces the time span in which relevant information is contained
and roughly doubles the relevant $\delta{E}_\mathrm{exc}$. For
example, to obtain a spectral resolution of 1 MeV, one needs to
simulate up to about 1200 fm/c.

Although excitation spectra are one of the most basic properties of the system, there
are many other dynamical features of interest. The multipole signals
in the time domain (printed in the protocol files) are as such
interesting quantities. One can have, e.g., a look at cross-talk
between the multipole channels. It is particularly interesting to
study excitation dynamics for varying excitation strength $\eta$, from
the regime of linear response (small $\eta$) deep into the non-linear
regime. It is inefficient to perform a full three-dimensional TDHF
calculation to obtain linear-regime excitation spectra for spherical nuclei. This is
better done in a dedicated RPA calculation on a spherical basis (see,
e.g.,~\cite{Rei92b}) for which there exist an overwhelming multitude
of codes. The realm of TDHF calculations of nuclear excitations are
spectra in deformed systems, stability analysis of exotic
configurations, and in particular non-linear dynamics.

There are many more details worth looking at.  One may check the
densities and currents to visualize the flow pattern associated with a
mode.  A most elaborate analysis deals with a phase-space picture of
nuclear dynamics by virtue of the Wigner transformation~\cite{Loe11a}.
The code allows saving all ingredients needed for such elaborate
analysis in dedicated output files, see Section~\ref{sec:silo}. It is
left to the user to work out the further steps of the analysis.

\subsubsection{Nuclear reactions}
\label{sec:react}

Collisions of nuclei are a prime application of nuclear TDHF. They were,
in fact, the major motivation for its realization \cite{Bon78a,Dav85a}.
The present code is designed to initialize such collision scenarios in
a most flexible manner. We start by explaining the simplest case of
a collision of two nuclei. First, we prepare the ground states of the
two nuclei as explained in Section~\ref{sec:statiter}. The static
solutions are centered around the origin $\vec{r}=0$ of their initial
grid.  The static wave functions $\psi_{\alpha,I}^{\mbox{(stat)}}$
where $I=1,2$ labels the two nuclei are shifted to new centers
$\vec{R}_I$ where the distance $|\vec{R}_2-\vec{R}_1|$ should be
sufficiently large that the nuclei have negligible overlap and
negligible Coulomb distortion from the other nucleus (the latter
condition usually only loosely fulfilled). The shifted wave functions
$\psi_{\alpha,I}(\vec{r}-\vec{R}_I,s)^{\mbox{(stat)}}$ are obtained by
interpolation on the grid. It is obvious that the collisions need a
larger numerical box than the static Hartree-Fock calculations. Thus, we may compute
the static wave functions on a smaller box since we are shifting and
interpolating the result anyway for dynamical initialization. At this point
we have the nuclei resting at a safe distance. To set them in motion
we need to give each nucleus a momentum
$\vec{P}_1=-\vec{P}_2$ (note that the total momentum of the combined
system still vanishes). Consequently, the initial configuration is given
by the Slater state built from the shifted and boosted single-particle
wave functions (see fragment initialization, point \ref{it:frag} in
Section~\ref{sec:init})
\begin{equation}
\begin{array}{ll}
   \psi_{\alpha,1}(\vec{r},s;t\!=\!0)
    =e^{\I\vec{p}_1\cdot\vec{r}}
    \psi_{\alpha,1}(\vec{r}-\vec{R}_1,s)^{\mbox{(stat)}},&
      \vec{p}_1 =\frac{\vec{P}_1}{A_1}
    \quad,
    \\
    \psi_{\alpha,2}(\vec{r},s;t\!=\!0)
    = e^{\I\vec{p}_2\cdot\vec{r}}
    \psi_{\alpha,2}(\vec{r}-\vec{R}_2,s)^{\mbox{(stat)}},&
    \vec{p}_2=\frac{\vec{P}_2}{A_2}
    \quad.
 \label{eq:inittdhf}
\end{array}
\end{equation}
The distance between the nuclei is large, but inevitably finite. This may
induce minor violations of orthonormality. Thus the
full set of wave functions (\ref{eq:inittdhf}) is orthonormalized as a final step of
initial preparation.

The occupation amplitudes $v_{\alpha,I}$ are taken over from the
static solution and frozen along the dynamical evolution. In fact,
most of the collision studies will be principally to explore
the dynamical features in the regimes of fusion and inelastic
collisions. It is then recommended to use the conceptually simplest
and most robust strategy, namely to start from simple Slater states
(not BCS states) for the two nuclei. This means we that in most cases
the static solution is calculated without pairing, fixing $v_\alpha=1$ and
including just as many states as there are nucleons.

The above example deals with two initial fragments. The code is more
flexible than that. It allows an initial state composed from several
fragments. The strategy remains the same as for the binary system. It
is simply repeated for each new fragment. 

The time evolution is performed as outlined in Section
\ref{sec:timevol}. It requires some effort to visualize the complex
dynamics which emerges in collisions. A rough picture is, again,
provided by the multipole moments. The quadrupole moment, e.g., can
serve as a measure of stretching of the total system. Small values
indicate a compound nucleus while asymptotically growing values signal
fragmentation. One may want, particularly in case of collisions, more
detailed pictures of the flow as, e.g., snapshots of the density of
current distributions and, ultimately, a full phase space picture
\cite{Loe11a}. Material for that can be output on demand, as detailed
in Section \ref{sec:silo}. Again, we leave it to the user to extract
the wanted information and to prepare it for visualization.

\subsection{Observables}
\label{sec:observables}

It was already mentioned in Sections~\ref{sec:collex} and
\ref{sec:react} which observables may be used to analyze nuclear
dynamics. We here briefly summarize the observables computed
and output in the code and indicate how further observables may be
extracted. Basic features of the description by the Skyrme
energy-density functional are, of course, energy and densities.  

\subsubsection{Multipole moments}
\label{sec:mulmom}

The gross features of the density distribution are well characterized
by the multipole moments. The most important moment is the
center of mass (c.m.)
\begin{subequations}
  \label{eq:mulmom}
  \begin{equation}
    \vec{R}^\mathrm{(type)}
    =
    \frac{\int \D^3r\,\vec{r}\,\rho^\mathrm{(type)}(\vec r)}{A}
    \quad,
    \label{eq:cmcart}
  \end{equation}
  where $A=\int \D^3r\,\rho(\vec r)$ is the total mass number and
  ``type'' can refer to proton c.m.\ from $\rho_p$, neutron c.m.\ 
  from $\rho_n$, isoscalar or total c.m.\  from the total density
  $\rho=\rho_p+\rho_n\equiv\rho_{T=0}$, or isovector moment related to
 the  isovector density $\rho_{T=1}=\frac{N}{A}\rho_p-\frac{Z}{A}\rho_n$.
  The
  definition of $\vec{R}^\mathrm{(type)}$ directly employs the
  Cartesian coordinate $r_i$. The same holds for the Cartesian
  quadrupole tensor
  \begin{equation}
    \mathcal{Q}_{kl}^\mathrm{(type)}
    =
    \int \D^3r \left(3(r_k-R_k)(r_l-R_l)-\delta_{kl}\sum_i(r_i-R_i)^2\right)
    \rho^\mathrm{(type)}(\vec r)
    \quad,
    \label{eq:Qcart}
  \end{equation}
  again for the various types as discussed above.  The matrix
  $\mathcal{Q}_{kl}$ is not invariant under rotations of the
  coordinate frame. There is a preferred coordinate system: the system
  of principle axes. It is obtained by diagonalizing
  $\mathcal{Q}_{kl}$. The quadrupole matrix in the principle-axis
  frame thus has only three non-vanishing entries $Q_{xx}$, $Q_{yy}$,
  and $Q_{zz}$ together with the trace condition
  $Q_{xx}+Q_{yy}+Q_{zz}=0$.  Although straightforward to define,
  higher Cartesian moments are messy and inconvenient to use.
  The spherical multipole moments are deduced from
  \begin{equation}
    Q_{lm}^\mathrm{(type)}
    =
    \int \D^3r\,r^lY_{lm}\,\rho^\mathrm{(type)}(\vec{r}+\vec{R})\;,
    \label{eq:spherQ}
  \end{equation}
  where $r=|\vec{r}|$ and $Y_{lm}$ are spherical harmonics.  The most
  important ones are monopole ($l=0$), dipole ($l=1$), and quadrupole
  ($l=2$) moments.  The latter are often expressed as dimensionless
  quadrupole moments
  \begin{equation}
    a_m
    =
    \frac{4\pi}{5}
    \frac{Q_{2m}}{AR^2}
    \quad,
    \label{eq:dimlessQ}
  \end{equation}
  with $R=r_0A^{1/3}$ a fixed radius derived from the total mass
  number $A$.
  This, again, could be defined for any ``type'', but is used, in
  practice, mainly for the isoscalar moments.  $r_\mathrm{rms}$ is
  the root-mean-square radius of the total density. The r.m.s. radii
  are defined as
  \begin{equation}
    r_\mathrm{rms}^\mathrm{(type)}
    =
    \sqrt{\frac{\int \D^3r\,(\vec{r}-\vec{R})^2\,\rho^\mathrm{(type)}(\vec{r})}
      {\int \D^3r\,\rho^\mathrm{(type)}(\vec{r})}}
    \label{eq:rmsrad}
  \end{equation}
  where ``type'' can be proton, neutron, or total. The isovector
  variant does not make sense here. 

 The dimensionless moments have
  the advantage of being free of an overall scale which was removed by
  the denominator $AR^2$. They allow characterization of the
  shape of the nucleus.  However, the general $a_m$ are not invariant
  under rotations of the coordinate frame. We obtain a unique
  characterization by transforming to the principle-axis system. These
  are defined by the conditions $a_{\pm 1}=0$ and $a_2=a_{-2}$. There
  remain only two shape parameters $a_0$ and $a_2$. These are often reexpressed
 as total deformation $\beta$ and triaxiality $\gamma$, often
  called Bohr-Mottelson parameters, through
  \begin{equation}
    \beta
    =
    \sqrt{a_0^2+2a_2^2}
    \quad,\quad
    \gamma
    =
    \mbox{atan}\left(\frac{\sqrt{2}\,a_2}{a_0}\right)
    \quad.
    \label{eq:triax}
  \end{equation}
  Triaxiality $\gamma$ is handled like an angle. It can, in principle,
  take all values between $0^o$ and $360^o$, but physically relevant
  parameters stay in the $0\ldots60^\circ$ range. The other sectors
  correspond to equivalent configurations~\cite{Gre05aB}.  We
  supply printouts of all the above variants of multipole moments to
  allow a most flexible analysis.
\end{subequations}

\subsubsection{Alternative way to evaluate the total energy}
\label{sec:alterenerg}

The key observable is the total energy $E_\mathrm{tot}$. It is computed as
given in Eqs. (\ref{eq:efunc}). More detailed energy observables are
provided by the s.p. energies (\ref{eq:spenerg}).  These can also be
used to compute the total energy. The traditional HF scheme deals with
pure two-body interactions and exploits that to simplify~\cite{Gre05aB}
\begin{equation}
  \label{eq:koopman}
  E_\mathrm{tot,HF}
  =
  \frac{1}{2}\sum_\alpha\left(t_\alpha+\epsilon_\alpha\right)
\end{equation}
where $t_\alpha=\langle\psi_\alpha|\hat{T}|\psi_\alpha\rangle$ is the
s.p. kinetic energy. This is possible because
\begin{equation*}
  \epsilon_\alpha=t_\alpha+u_\alpha
  \quad,\quad
  u_\alpha
  =
  \sum_{\beta}\left[v_{\alpha\beta\alpha\beta}-v_{\alpha\beta\beta\alpha}\right]
  -\tfrac1{2}   v_\alpha=\epsilon_\alpha-t_\alpha
  \quad,
\end{equation*}
where $u_\alpha$ is the s.p. mean-field potential energy and $v$ the
two-body interaction, and
\begin{equation*}
  E_\mathrm{tot,HF}
  =
  \sum_\alpha t_\alpha
  +
  \frac{1}{2}\sum_{\alpha\beta}
  \left[v_{\alpha\beta\alpha\beta}-v_{\alpha\beta\beta\alpha}\right]
  \quad.
\end{equation*}
The Skyrme force does not simply have this two-body structure. Still
the total energy is very often computed along the strategy of
Eq.~ (\ref{eq:koopman}). However, the density dependence requires
augmenting this recipe by a rearrangement energy which accounts for a
contribution missing in the simple recipe (\ref{eq:koopman}).  The
extension to the Skyrme energy thus reads 
\begin{subequations}
  \label{eq:koopman2}
  \begin{eqnarray}
    E_\mathrm{tot,HF}
    &=&
    \frac{1}{2}\sum_\alpha\left(t_\alpha+\epsilon_\alpha\right)
    +  E_{3,\rm corr} + E_\mathrm{C,corr}
    \quad,
    \\
    E_{3,\rm corr} 
    & = & 
    \int \D^3r \frac{\alpha}{6}  \rho^{\alpha}
    \left[b_3\rho^2 - b'_3 (\rho_{p}^2+\rho_{n}^2) \right]
    \quad,
    \label{eq:e3corr}
    \\
    E_\mathrm{C,corr}
    &=&  \frac{1}{4} \left( \frac{3}{\pi} \right)^{1/3}
    \int \D^3r\, \rho_{pr}^{4/3}
    \quad.
    \label{eq:ecorc}
  \end{eqnarray}
\end{subequations}
In the code the total energy is computed both ways, from the
straightforward Skyrme energy (\ref{eq:efunc}) as well as from the
above recipe (\ref{eq:koopman2}). Numerically these values are close but not
identical.

\subsection{Discretization}
\subsubsection{Data types}
In the module {\tt Params} a type {\tt db} is defined for 12-digit
accuracy and on all present machines should amount to double
precision, i.~e., {\tt REAL(db)} and {\tt COMPLEX(db)} are actually
{\tt REAL(8)} and {\tt COMPLEX(8)} for most compilers. Keeping the symbolic type
throughout of course makes the code more flexible for future hardware.
Using single precision is not recommended.

Since the external libraries are based on C or Fortran-77 coding,
special care has to be taken in this respect. The FFTW3 library stores
its plans in 8-byte integers, which in the modules {\tt Coulomb} and
{\tt Fourier} are defined using the type {\tt C\_LONG} from the
system-supplied module {\tt ISO\_C\_BINDING}. If this is not
available, they may be defined as {\tt INTEGER(8)} or if that also
causes problems, {\tt DOUBLE PRECISION}, which certainly corresponds
to at least 8 bytes. For the {\tt LAPACK} routines, double-precision
real and complex variables are necessary, which we also define using
``{\tt db''}. If {\tt db} should ever be changed in such a way that
{\tt REAL(db)} no longer corresponds to 8 bytes, different {\tt LAPACK}
routines must be selected.

Note that data conversion needs some care. If {\tt a} and {\tt b} are
double precision, {\tt CMPLX(a,b)} is not; according to the standard
it returns the default accuracy, which on many machines will still be
single precision. Thus that for example in {\tt EXP(CMPLX(a,b))} the
exponential would be evaluated in single precision. That is why in the
code the expression {\tt CMPLX(a,b,db)} is used consistently; this is
also safe for future changes in accuracy.

The other conversion functions used are safe. {\tt AIMAG} is generic
and {\tt REAL} reproduces the accuracy of (only) a {\em complex
  argument}.

\subsubsection{Grid definition}
\label{sec:griddef}
All wave functions and fields are defined on a three-dimensional
regular Cartesian grid of {\tt nx} by {\tt ny} by {\tt nz} grid
points. {\bf {\tt nx}, {\tt ny}, and {\tt nz} must be even numbers}.
The physical spacing between the points is given as {\tt dx}, {\tt
  dy}, and {\tt dz} (in fm). In principle these could be different for
the three directions, but since this will lead to a loss of accuracy
it is highly recommended to give the same value to all of them. A
typical range is 0.5--1.0~fm.

The grid is automatically arranged in such a way that in each
direction the same number of grid points are located on both sides of the
origin. This means that the three-dimensional origin is in the center
of a cubic cell and has the advantage that exact parity properties for
the wave functions can be maintained. The coordinate values for e.~g.,
the $x$-direction are thus:
\begin{equation}\begin{split}
 -\frac{{\tt nx}-1}{2}&{\tt dx},\quad -\frac{{\tt nx}-1}{2}{\tt
  dx}+{\tt dx},\quad\ldots \quad -\frac{{\tt dx}}{2},\\
  &+\frac{{\tt  dx}}{2},\quad \ldots\quad \frac{{\tt nx}-1}{2}{\tt
    dx}. 
\end{split}
\end{equation}
The corresponding values are available in the arrays {\tt x(nx)}, {\tt
  y(ny)}, and {\tt z(nz)}.

\subsubsection{Derivatives}
\label{sec:deriv}

The computation of the kinetic densities and currents and the
application of the mean-field Hamiltonian require first and second
derivatives at several places in the code. We define them in Fourier
space. For simplicity, the strategy is explained here for one
dimension. The generalization to 3D is obvious.

The {\tt nx} discrete grid points $x_\nu$ in coordinate space are
related to the same number of grid points $k_n$ in Fourier space
(physically equivalent to momentum space) as
\begin{subequations}
\label{eq:FT}
\begin{eqnarray}
  x_\nu
  &=&
  \left(-\frac{{\tt nx}-1}{2}+\nu\right){\tt dx}
  \quad,
  \nu=1,...,{\tt nx}
  \quad,
\\ \label{eq:kvalues}
  k_n  &=&
  (n-1) {\tt dk},\quad n=1,\ldots {\tt nx}/2 \quad,
\\ \nonumber
  k_n&=&(n-{\tt nx}-1)\,{\tt dk},\quad n={\tt nx}/2+1,\ldots,{\tt nx} \quad,
\\
  {\tt dk}
  &=&
  \frac{2\pi}{{\tt nx}\cdot{\tt dx}}
  \quad.
\end{eqnarray}
Note the particular indexing for the $k$-values. In principle, the
values $k_n=(n-1){\tt dk}$ for all $n$ are equivalent for the Fourier
transform, but for the second half of this range the negative
$k$-values should be chosen because of their smaller magnitude. For
the Fourier expansion, $k=-{\tt dk}$ and $k=({\tt nx}-1){\tt dk}$ are
equivalent because of periodicity in $k$-space.

A function $f(x_\nu)$ in coordinate space is connected to a
function $\tilde{f}(k_n)$ in Fourier space by
\begin{eqnarray}
  \tilde{f}(k_n)
  &=&
  \sum_{\nu=1}^{\tt nx}
  \exp{\left(-\I k_nx_\nu\right)}f(x_\nu) 
  \quad,
\label{eq:FTforward}\\
  f(x_\nu) 
  &=&
  \frac{1}{{\tt nx}}\sum_{n=1}^{\tt nx}
  \exp{\left(\I k_nx_\nu\right)}\tilde{f}(k_n)
\label{eq:FTbackward}
\end{eqnarray}
\end{subequations}
This complex Fourier representation implies that the function $f$ is
periodic with $f(x+{\tt dx}\cdot{\tt nx})=f(x)$. The appropriate
integration scheme is the trapezoidal rule which complies with the above
summations adding up all terms with equal weight.
The derivatives of the exponential basis functions are
\begin{equation}
  \frac{\D^m}{\D x^m}\exp{(\I k_nx)}
  =
  (\I k_n)^m\exp{(\I k_nx)}
  \quad.
\end{equation}
Computation of the $m$th derivative thus becomes a trivial
multiplication by $(\I k_n)^m$ in Fourier space. Time critical
derivatives are best evaluated in Fourier space using the fast Fourier
transformation (FFT). To that end a forward transform
(\ref{eq:FTforward}) is performed, then the values $\tilde{f}(k_n)$
are multiplied by $(\I k_n)^m$ as given in
Eq.~(\ref{eq:kvalues}) and finally transformed $(\I
k_n)^m\tilde{f}(k_n)$ back to coordinate space by the transformation
(\ref{eq:FTbackward}). This strategy is coded in the subroutines
\texttt{cdervx}, \texttt{cdervy}, and \texttt{cdervz} contained in
module \texttt{Levels}.  It is used for derivatives of wave functions
provided the switch {\tt TFFT} is set.

For coding purposes, it is often useful to perform derivatives as a
matrix operation directly in coordinate space. The derivative matrices
are built by evaluating the double summation of forward and backward
transform ahead of time. For the $m$th derivative this reads
\begin{equation*}\begin{split}
  f^{(m)}(x_\nu)
  &=
  \frac{1}{\tt nx}
  \sum_n\exp{(ik_nx_\nu)}(\I k_n)^m
  \sum_{\nu'}\exp{(-\I k_nx_{\nu'})}f(x_{\nu'})
\\
  &=
  \sum_{\nu'}\;
  \underbrace{
  \frac{1}{\tt nx}
  \sum_n\exp{(\I k_nx_\nu)}(\I k_n)^m
  \exp{(-\I k_nx_{\nu'})}}_{D^{(m)}_{\nu\nu'}}\;
  f(x_{\nu'})
  \quad.
\end{split}
\end{equation*}
From here, the detailed handling depends on the order of
derivative. The $k_n$ run over the values
$k_n=0,
\pm{\tt dk},\pm 2{\tt dk},...,({\tt nx}/2-1){\tt dk},+{\tt dk}\,{\tt
  nx}/2$.
Here the index ordering given in Eq.~(\ref{eq:kvalues}) does not matter
as the index is summed over.
Note that the first and the last value come alone while all others
come in pairs of $\pm$ partners. These pairwise terms can be combined
into a sine function for $n=1$ and a cosine for $n=2$. The derivative
matrices thus read in detail
\begin{subequations}
\begin{align}
\label{eq:derv1}
  D^{(1)}_{\nu\nu'}
  =&
  -\frac{2{\tt dk}}{\tt nx}
  \sum_{n=1}^{{\tt nx}/2-1}n\sin(k_n{\tt dx}(\nu\!-\!\nu'))
\notag\\
  &
  -\frac{{\tt dk}}{\tt nx}
  \frac{{\tt nx}}{2}\sin((\nu\!-\!\nu'){\tt dk}\,{\tt nx}/2)
  \quad,
\\  
\label{eq:derv2}
  D^{(2)}_{\nu\nu'}
  =&
  -\frac{2{\tt dk}^2}{\tt nx}
  \sum_{n=1}^{{\tt nx}/2-1}n^2\cos(k_n{\tt dx}(\nu\!-\!\nu'))
\notag\\
  &
  -\frac{{\tt dk}^2}{\tt nx}
  \left(\frac{{\tt nx}}{2}\right)^2\cos((\nu\!-\!\nu'){\tt dk}\,{\tt nx}/2)
  \quad.
\end{align}
\end{subequations}
A word is in order about the first derivative. The upper point in the
$k$-grid, ${\tt dk}\,{\tt nx}/2$, is ambiguous. Exploiting periodicity,
it could be equally well $-{\tt dk}\,{\tt nx}/2$. In order, to deal
with a $\pm k$ symmetric derivative we have anti-symmetrized this last
point. The price for this is a slight violation of hermiticity which,
however, should be very small as we anyway should not have significant
wave-function contributions at the upper edge of the $k$-grid.

The derivative matrices $D^{(m)}_{\nu\nu'}$ can be prepared ahead of
time and are then at disposal for any derivative in the course of the
program. Actually, the matrices for the first derivative are prepared
in routine \texttt{sder} and for the second derivative in routine
\texttt{sder2}, both contained in module \texttt{Grids}. These
routines are applied to generate the derivative matrices
\texttt{derv1x}, \texttt{derv1y}, \texttt{derv1z}, \texttt{derv2x},
\texttt{derv2y}, and \texttt{derv2z}, for first and second derivatives
in the $x$, $y$ and $z$ directions.

The matrix formulation of the derivatives is used in the code in two
ways: on the one hand, the derivatives of the real-valued densities,
currents, and mean-field components are always calculated using these
derivative matrices, because they are real and the Fourier transform
method would require converting them to complex values (using special
Fourier techniques for real arrays is in principle possible but has
not been worked out yet). In addition the user can switch to using the
matrix method everywhere, which may give a slight speed advantage for
small grid dimensions.

\subsubsection{Boundary conditions}
The code uses a periodic Fourier transform to calculates derivatives.
This is valid only with {\em periodic boundary conditions}. Thus in
principle the wave functions and potentials are assumed to be repeated
periodically in each Cartesian direction. Because of the short range
of the nuclear force, this is not a serious problem in most cases; at
higher energies, however, the emission of low-density material from
the nuclei can interfere with the dynamics in the neighboring box and
cause problems in the conservation of energy and angular momentum; for
a detailed discussion see~\cite{Guo08a}.  This is aggravated by
  the fact that even with periodic boundary conditions periodicity is
  truly fulfilled only for the wave functions and mean-field
  components. Since the vector $\vec r$ itself is not periodic but
  jumps at the boundary, operators such as the orbital angular
  momentum are not periodic.

  On the other hand, for the Coulomb field with its long range this
  would be clearly wrong. Therefore a computation of the Coulomb
  potential for the boundary condition of an isolated charge
  distribution is implemented in addition to the periodic one, see the
  manual for the module {\tt Coulomb}. This is selected by the logical
  input variable {\tt periodic} {\bf and applies only to the Coulomb
    potential}.

\subsubsection{Wave function storage}
Module {\tt Levels} handles the single-particle wave functions and
associated quantities.

The principal array for the wave function is called {\tt psi}, which
is of type {\tt COMPLEX(db)}. Its dimension is {\tt
  (nx,ny,nz,2,nstloc)}, where the first three indices naturally refer
to the spatial position. The 4th index corresponds to spin: index~1
refers to spin up and~2 to spin down, quantization being along the
$z$-direction.

The last index numbers the wave functions. If the code is run on a
single node, the value is {\tt nstmax}, the total number of
single-particle wave functions. They are divided up into neutron and
proton states, with the index range given by {\tt npmin} and {\tt
  npsi}. The sub-ranges are:
\begin{itemize}
\item {\tt npmin(1)}$\ldots${\tt npsi(1)} : the neutron states,
\item {\tt npmin(2)}$\ldots${\tt npsi(2)} : the proton states.
\end{itemize}
In the present code {\tt npmin(1)=1} and {\tt npsi(2)=nstmax}.

\begin{figure}
\centerline{\includegraphics[width=0.7\linewidth]{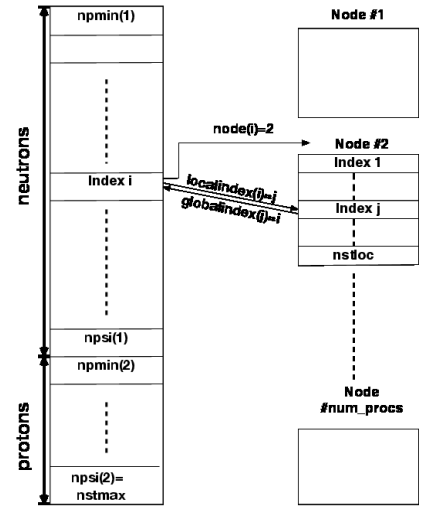}}
\caption{
Storage arrangement of the single-particle wave functions. On the left
the case for single-processor or {\tt OpenMP} is shown, which for the
case of distributed memory under {\tt MPI} is mapped to the individual
processors as shown on the right. Note that each node will have its
own values of {\tt nstloc} and {\tt globalindex}.
}
\end{figure}

If the code is run in parallel (MPI) on several nodes, only {\tt nstloc}
single-particle wave functions are stored on a given node, where {\tt
  nstloc} may vary. Pointers are then defined to indicate the
relationship between the local index and that in the global array of
wave functions. For details see the section on parallelization.

There are a number of arrays containing the physical properties of the wave
functions, such as the single-particle energy. The names start with
{\tt sp\_} and they are defined in module {\tt Levels}. They are not
split up in the parallel case, but on each node only the pertinent
index positions are used.

\subsubsection{Densities and currents}
The various densities necessary for constructing the mean field are
actually kept in separate arrays and can be output onto data files for
later analysis (see subroutine {\tt write\_densities}). The
dimensioning is {\tt (nx,ny,nz,2)} with the last index referring to
isospin for scalar densities, so that {\tt rho(:,:,:,1)} is the
neutron density and {\tt rho(:,:,:,2)} the proton density. For vector
densities there is an additional index with values~1 to~3 for the
Cartesian direction, thus {\tt sdens(nx,ny,nz,3,2)} containing the
spin density in each direction for neutrons and protons.

Since it is often not necessary to keep the neutron and proton
contributions separate, subroutine {\tt write\_densities} has the option of
adding them up before output.

\subsection{Initialization}
\label{sec:init}
A particular strength of the code is its flexible initialization.
There are essentially three types of initialization, which can be
selected through the input variable {\tt nof}:

\begin{enumerate}
\item\label{it:iniho} {\bf Harmonic oscillator:} {\tt nof=0}: this is
  applicable only to static calculations. The initial wave functions
  are generated from harmonic oscillator states with initial radii
  {\tt radinx}, {\tt radiny}, and {\tt radinz} in the three
  directions. It is advisable to choose the three radii different to
  avoid being kept in a symmetric configuration for non-spherical
  nuclei. Note that this is a very simple initialization and has some
  defects; for example, the initial deformation is controlled more by
  the occupation of the oscillator states than by the radius
  parameters. This should eventually be replaced by, e.~g., Nilsson
  wave functions.

  For this case the type of nucleus is determined by the input numbers
  {\tt nneut} and {\tt nprot} giving the number of neutrons and
  protons, while {\tt npsi} can be used to add some unoccupied states
  (this sometimes leads to faster convergence).

\item\label{it:frag} {\bf Fragment initialization:} {\tt nof>0}: wave
  functions for a number {\tt nof} of fragments are read in and
  positioned in the grid at certain positions. The wave functions are
  read from files produced by the static code with the file names
  given by the input {\tt filename}, they are positioned at
  center-of-mass positions {\tt fcent} and given am initial velocity
  controlled by {\tt fboost}. The code determines the number of wave
  functions needed from these data files and also checks the agreement of
  Skyrme force and grid used. This initialization is used, e.g., for
  nuclear reactions (see Section~\ref{sec:react}).

  The number of fragments read in is arbitrary, but there are two
  special cases:
  \begin{itemize}
  \item for {\tt nof=1} a single fragment is read in. This can be
    useful for initializing with static wave functions to study
    collective vibrations in a nucleus using the TDHF mode.
  \item for {\tt nof=2} a special initialization can be done where the
    initial velocities are not given directly but computed from a
    center-of-mass energy {\tt ecm} and an impact parameter {\tt b}.
  \end{itemize}
  
\item {\bf User initialization:} a user-supplied routine {\tt
    user\_init} can be employed to set up the wave functions in any
  desired way. The only condition is that the index ranges etc.\ are
  set up correctly and the wave function array {\tt psi} is filled
  with the proper values. It was found useful, e.~g., to use initial
  Gaussians distributed in various geometric patterns for
  $\alpha$-cluster studies.
\end{enumerate}

\subsection{Restarting a calculation}\label{sec:restart}
Sometimes it is necessary to continue a calculation that was not run
to the desired completion because of a machine failure or because the
number of iterations or time steps was set too low. In such cases the
last wave function file with name {\tt wffile}, which is generated at
regular intervals of {\tt mrest} iterations or time steps, can be used
to initialize a continuation. The program handles this in a simple fashion: if
the logical variable {\tt trestart} is input as {\tt TRUE}, it sets up
an initialization with one fragment (read from the initialization file)
placed at the origin and with zero
velocity. The only other modifications to the regular setup are then
to take the initial iteration number and time from that file instead
of starting at zero, as well as suppressing some unneeded initialization steps.

This flexible restart makes it possible to use a different grid for
the continuation in the sense that the grid spacings must agree, but
the new grid can be larger than the old one.

\subsection{Accuracy considerations}\label{sec:accuracy}
The grid representation and solution methods introduced above depend
on several numerical parameters. Their proper choice is crucial
for the accuracy and speed of the calculations. In this Section, we
want to briefly address the dependence on numerical parameters. An
extensive discussion of grid representations and static iteration is
found in~\cite{Blu92a}.

\begin{figure}[h]
\centerline{\includegraphics[width=\linewidth]{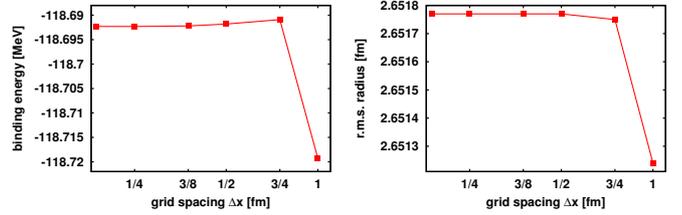}}
\caption{\label{fig:static-grid} Binding energy (left) and
  r.m.s. radius (right) of $^{16}$O computed for the force SkI3 drawn
  as functions of grid spacing $\Delta x=\Delta y=\Delta z$. A
  logarithmic scale us used for $\Delta x$. The number of grid points
  has been chosen to keep the box size constant at $N_x\Delta x=24$
  fm.}
\end{figure}
Figure \ref{fig:static-grid} shows the sensitivity with respect to the grid
spacings $\Delta x$, $\Delta y$, $\Delta z$. The trend is the same for
both observables, energy and radius: The results have very high
quality and change very little up to $\Delta x=$ 0.75 fm. They quickly
degrade above that spacing. But even at $\Delta x=$ 1 fm, we still
find an acceptable quality which suffices for most applications,
particularly for large scale explorations. If high accuracy matters, 
$\Delta x\approx$ 0.75 fm should be chosen;  not much is gained 
by going to even finer gridding. This holds for
ground states and moderate excitations. High excitations and fast
collisions may require a finer mesh. Note that the maximum
representable kinetic energy is
$E_\mathrm{kin,max}=(\hbar^2/2m)(\pi/\Delta x)^2$, which amounts to
about 200 MeV for $\Delta x=$ 1 fm. The actual energies of interest
should stay far below this limit. It is an instructive exercise to
study uniform center-of-mass motion at various velocities to explore the
limits of a given representation.

The number of grid points $N_x=N_y=N_z$ in the tests of Figure
\ref{fig:static-grid} were chosen such that the box size was the
same in all cases.  The actual choice of $N_x$ depends sensitively on
the system, its size and separation energy. As a rule of thumb, the
density decreases asymptotically as
$\rho\propto\exp{(-2\sqrt{2m\varepsilon_N}r/\hbar)}$ where
$\varepsilon_N$ is the single particle energy of the least bound
state. One should aim for at least $\rho<10^{-8}$ at the boundaries.

\begin{figure}
\centerline{\includegraphics[width=0.5\linewidth]{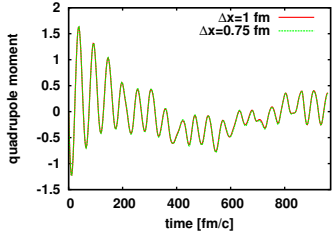}}
\caption{\label{fig:dynamic-grids} Time evolution of the quadrupole
  momentum for two different grid spacings $\Delta x=\Delta y=\Delta
  z$ and constant box size of 24 fm. The test case is $^{16}$O excited by
  an instantaneous boost computed with the force SkI3.}
\end{figure}
Figure \ref{fig:dynamic-grids} explores the effect of grid spacing for
dynamics. Two different spacings are compared for a quadrupole
oscillation following an instantaneous quadrupole boost. Practically
no difference can be seen for the ``safe choice'' $\Delta x=$ 0.75 fm
and the robust choice $\Delta x=$ 1 fm. Dynamical applications,
oscillations and collisions, are in general less demanding and can
be performed very well with $\Delta x=$ 1 fm. This is pleasing as dynamical
calculations are usually much more costly than purely static ones.

There are two parameters regulating the static iteration according to
Eq.~(\ref{eq:dampstep}), the damping energy {\tt e0inv} and the step
size {\tt x0dmp}. {\tt e0inv} should correspond to the depth of
the binding potential. The overall step size {\tt x0dmp} can be of
order of one if {\tt e0inv} is well chosen. Nuclear binding is very
similar all over the chart of nuclei. This allows to develop one safe
choice for nearly all cases. We recommend {\tt e0inv}$\approx 100$ MeV
together with {\tt x0dmp}$\approx 1/2$, reducing the latter slightly
if convergence problems appear. Of course,
 a few percent in iteration speed my be gained by fine-tuning these
parameters for a given case, but this is not worth the effort unless
large scale surveys for a given class of nuclei and forces are
planned.

\begin{figure}
\centerline{\includegraphics*[width=0.85\linewidth]{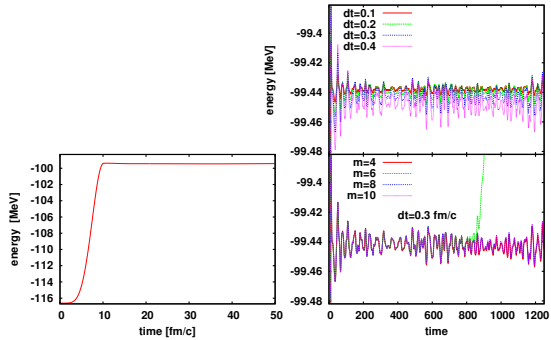}}
\caption{\label{fig:dynamicSV-energ} Time evolution of the binding energy
  of $^{16}$O after an excitation by a soft sin$^2$ pulse of width 20
  fm/c computed for the force SV-bas~\cite{Klu09a} with a grid spacing
  $\Delta x=0.75$ fm and box size of 24 fm. Results are shown for
  different sizes of time step {\tt dt} and different order of Taylor
  expansion {\tt m} of the exponential evolution. The left panel shows
  the full evolution from ground state energy to the excited
  energy. The right panels concentrate on the times after the pulse is
  over in a narrower energy range relevant for this excitation.}
\end{figure}

The time stepping using the exponential propagator has the two parameters,
step size {\tt dt} and order {\tt mxp}$=m$ of the Taylor expansion
(\ref{eq:timepower}) of the exponential. Intuitively, one expects that
small {\tt dt} and large {\tt mxp} improve the quality of the
step. An efficient stepping scheme, however, looks for the largest {\tt
  dt} and smallest {\tt mxp} which still provide acceptable and
stable results. It is hard to give general rules as good working
values for the parameters depend on all details of the actual calculation:
gridding, nuclei involved, excitation energy, and kind of
excitation. 

Figure \ref{fig:dynamicSV-energ} demonstrates the dependence of a
typical dynamical evolution on these time-stepping parameters. We
consider a time interval up to 1260 fm/c which is a long time for
heavy-ion collisions and just sufficient for a spectral analysis of
oscillations~\cite{Rei07a}. The excitation is done by a soft sin$^2$
pulse of finite extension in time. The energy increases during the
initial excitation phase, as can be seen from the left panel in the
figure. After the external pulse is over, energy conservation holds,
which is nicely seen at plotting resolution in the left
panel. Normalization should be conserved at all times.  Both
conservation laws serve as tests for the time step. Norm is conserved
up to at least six digits for all cases and times shown in Figure
\ref{fig:dynamicSV-energ}. The energy is more critical. The right
panels show the energy in a small window around the final energy after
the excitation phase is over. The right lower panel shows a variation
of the Taylor order $m$ for fixed time step. The most prominent effect
is the sudden turn to catastrophic failure for $m=6$. In fact,
propagation by approximate exponential evolution explodes sooner or
later in all cases. The art is to extend the stable interval by
a proper choice of the stepping parameters. It is plausible that the
cases with $m>6$ maintain stability longer because the exponential is
better approximated. It is surprising that $m=4$ is also stable over
the whole time interval. There seem to be subtle cancellations of
error going on. Considering the stable signals, we see very little
differences between the cases. One may generally be happy with low
$m$. It is mainly stability demands which could call for larger
$m$. Note that this is not a generic result.  Stability for a given
test case should be checked once in a while and particularly before
launching larger surveys.

The right upper panel in Figure \ref{fig:dynamicSV-energ} shows
results for different {\tt dt} (as we have seen, the $m$ values are
not important as long as we achieve stable results). Here we see a
clear dependence on the step size. The energies remain constant in the
average. But there are energy fluctuations and these depend
sensitively on {\tt dt}. Smaller {\tt dt} yields smaller fluctuations.
As far as one can read off from the figure, the amplitude of the
fluctuations shrink $\propto${\tt dt}$^2$. It depends on the intended
analysis to which level of precision the time
evolution should be driven. A value of {\tt dt}$\approx$ 0.4 fm/c will be acceptable in most
cases because the average trend remains far smaller than the
fluctuations. Here also it must be emphasized that this is not a
generic number. Forces with lower effective mass (SV-bas has
$m^*/m=0.9$) are more demanding and usually require smaller {\tt
  dt}. On the other hand, running propagation without the spin-orbit term
allows even larger time steps because the spin-orbit potential is the
most critical piece in the mean-field Hamiltonian. The mix of
$\mathbf{p}$ and $\mathbf{r}$ imposes high demands on the numerical
representations. We again strongly recommend running a few
tests when switching forces or excitation schemes.

\section{Code structure}
The code is completely modularized to provide as large a degree of
encapsulation as possible in order to ease modification. Most modules
read their operating parameters from an associated {\tt NAMELIST} and
have their local initialization routines. In addition, a modern style
of programming is used that employs a minimum number of local
variables and streamlined array calculations that make the code lines
very close to the physical equations being solved.

Here we give a brief overview over the modules and their
purpose. There is a comprehensive manual supplied with the electronic
version that gives a detzailes description of all modules. The source
files containing the modules have the same names, but all in lower
case, with an extension of {\tt .f90}. The higher-level modules are:

\begin{description}
\item{Main program}: It calls initialization routines, sets up the
  initial wave functions using either harmonic oscillator states or
  reading wave functions of static Hartree-Fock solutions from given
  input files (module {\tt Fragments}). It then calls either {\tt
    statichf} from module {\tt Static} or {\tt dynamichf} from module
  {\tt Dynamic} to run the calculation.
\item{Static}: This contains the code for the static iterations {\tt
    statichf} and the subroutine {\tt sinfo} to generate output of the
  results.
\item{Dynamic}: runs the dynamic calculation in {\tt dynamichf} and
  generates output in {\tt tinfo}. Also controls the inclusion of an
  external excitation implemented in module {\tt External}.
\item{Densities}: calculates the densities and current densities by
  summing over the single-particle states.
\item{Meanfield}: contains the central physics calculation: the
  computation of the components of the mean field (subroutine {\tt
    skyrme}) and the application of the single-particle Hamiltonian to
  a wave function (subroutine {\tt hpsi}).
\item{Coulomb}: calculation of the Coulomb potential.
\item{Energies}: calculation of the total energies and its various
  contributions. 
\item{External}: calculation of the action of an external potential or
  initial collective boost of the wave functions.
\item{Pairs}: Implementation of the pairing correlations in the BCS
  approximation.
\item{Moment}: calculation of moments and deformation parameters for
  the bulk density.
\item{Twobody}: attempts to divide up the system into two separated
  nuclei and to calculate their properties and relative motion.
\end{description}

The lower-level supporting modules are:
\begin{description}
\item{Params}: general parameters used throughout the code. 
\item{Forces}: defines parameters of the Skyrme force and the pairing
  interaction and constructs them according to input.
\item{Grids}: defines everything associated with the numerical grid
  and sets it up.
\item{Levels}: definition of the single-particle wave functions and
  elementary operations on them such as derivatives.
\item{Fragments}: controls the reading of static wave functions from
  precomputed data and setting them up in the grid.
\item{Inout}: contains the subroutines for I/O of wave functions and
  densities. 
\item{Trivial}: defines some very basic operations on wave functions
  and densities.
\item{Fourier}: sets up the transform plans for the {\tt FFTW3}
  package to calculate Fourier transforms of wave functions and
  densities.
\item{Parallel}: This comes in two versions. The source file {\tt
    parallel.f90} contains the routines to handle {\tt MPI} message
  passing, while {\tt sequential.f90} sets up essentially dummy
  replacements for sequential or {\tt OpenMP} mode.
\item{User}: contains a sample user initialization code which can be
  used as a template for more complicated setups.
\end{description}

\section{Parallelization}
For both OpenMP~\cite{Cha08aB} and MPI~\cite{none12} the code can be
run in parallel mode. Parallelization for the static mode works in
OpenMP but not in MPI: the reason is in the orthogonalization step
which is not easily amenable for distributed-memory parallel
computation. This should be worked on in the future.

MPI and OpenMP can be used jointly if there are computing nodes with
multiple processors.

In both cases parallelization is done over the wave functions. The
code applies the time-development operator or the gradient iteration, which use the fixed set
of mean-field components, to each wave function, and this can naturally be
parallelized. Computing the mean fields and densities by summing up
over single-particle wave functions is also easily parallelizable. 

The library {\tt FFTW3}~\cite{Fri05a} itself can also run on multiple
processors in parallel. This can be used in addition to OpenMP or MPI,
but was found to be helpful only in the sequential version of the
code.

\subsection{OpenMP}
The application of the subroutine {\tt tstep} for propagating one wave
function for one time step, and of {\tt add\_densities} for adding one
wave function's contribution to the mean fields is done in
parallel loops. The only complicating factor is that the densities,
being accumulated in several subsets, must be kept separate using the
{\tt REDUCTION(+)} clause of OpenMP. The summation cannot be done
internally in {\tt add\_densities}, because for the half time step the
wave functions are immediately discarded after adding their
contribution to the densities to avoid having to store the full set at
half time. Thus only the combined {\tt tstep}-{\tt add\_densities}
loop should be parallelized.

The OpenMP program version can be compiled using the appropriate
compiler option. A separate {\tt Makefile.openmp} is provided which
just contains the {\tt -fopenmp} option for the GNU compiler.  The
number of parallel threads is not set by the code: the user should set
the environment variable {\tt OMP\_NUM\_THREADS} to the desired number.

\subsection{MPI}
In principle MPI uses the same technique as OpenMP, parallelizing over
wave functions. In this case, however, each node contains only a
fraction of the wave functions. This has several consequences:
\begin{enumerate}
\item The time-stepping of the wave functions can be done
  independently on each node, but requires that the densities are
  broadcast to all nodes after each half or full time step by summing
  up partial densities from the nodes in subroutine {\tt
    collect\_densities}.
\item The other calculation that uses wave functions directly is that
  of the single-particle properties. These are calculated on each node
  for the wave functions present on that node and then collected using
  subroutine {\tt collect\_sp\_properties}.
\item Only one node must be allowed to produce output. This is
  regulated by choosing node \#0 and setting the flag {\tt wflag}.
\item The saving of the wave functions is done in the following way:
  Node zero writes a header file containing the job information on
  file {\tt wffile}, then each node writes a separate file {\tt
    wwfile.001}, {\tt wwfile.002}, and so on up to the number of
  nodes. This avoids having to collect the wave functions on one node.
\item Using these parallel output files as fragment initialization or
  restart files is handled so flexibly that they can be read into a
  different nodal configuration or even a sequential run.
\end{enumerate}

The MPI version needs the appropriate compiler and linker calls for
the system used. The sequential or OpenMP versions are obtained simply
by linking with {\tt sequential.f90} instead of {\tt parallel.f90},
which replaces the MPI calls with a set of dummy routines and sets up
the descriptor arrays for the wave function allocation in a trivial
way.

The {\tt Makefile.mpi} shows the procedure; in practice systems differ
considerably and the user should look up the compilation commands for
his particular system.

\section{Input description}
All the input is through {\tt NAMELIST} and many variables have
default values. {\bf The {\tt NAMELIST}s should be in this file in the
  order in which they are described here, any {\tt NAMELIST} not used
  for a particular job may be omitted or left in the input file, in
  which case it is ignored}. The input is from standard input, so if
the data are prepared in a file {\tt inputdata} and the large output
listing is to go into {\tt output}, the code should be
run using, e.~g.,
\begin{verbatim*}
./sky3d.seq < inputdata > output
\end{verbatim*}

\subsection{Namelist  {\tt files}}\texttt{files}
\label{sec:filenamein}
This {\tt NAMELIST} contains names for the files used in the code.
They are defined in module \texttt{Params} and are:
\begin{description}
\item{\texttt{wffile}:} file to contain the static single-particle
  wave functions plus some additional data. This can be used for
  fragment initialization or for restarting a job. Default: {\tt
    'none'}, i.~e., nothing is written.
\item{\texttt{converfile}:} contains convergence information for the
  static calculation. Default: {\tt conver.res}.
\item{\texttt{monopolesfile}:} contains moment values of monopole
  type. Default: {\tt monopoles.res}.
\item{\texttt{dipolesfile}:} contains moment values of dipole type.
  Default: {\tt dipoles.res}.
\item{\texttt{quadrupolesfile}:} contains moment values of
  quadrupole type. Default: {\tt quadrupoles.res}.
\item{\texttt{momentafile}:} contains components of the total
  momentum. Default: {\tt momenta.res}.
\item{\texttt{energiesfile}:} energy data for time-dependent calculations.
  Default: {\tt energies.res}.
\item{\texttt{spinfile}:} time-dependent total, orbital, and spin
  angular-momentum data as three-dimensional vectors.
\item{\texttt{extfieldfile}:} 
time dependence of expectation value of the external field.
\end{description}

\subsection{Namelist {\tt force}} \texttt{force}
This defines the Skyrme force to be used. In most cases it should just
use two input values:
\begin{description}
\item[{\tt name}]: the name of the force, referring to the predefined
  forces in {\tt forces.data}.
\item[{\tt pairing}]: the type of pairing, at present either {\tt
    NONE} for no pairing,  \texttt{VDI} for the volume-delta
  pairing, or \texttt{DDDI} for density-dependent delta pairing. 
  The pairing parameters are included in the force definition. {\bf Note
  that the pairing type must be written in upper case.}
\end{description}
There is also the possibility for inputting a user-defined force; this
is described in detail with module \texttt{Forces} in the online
technical documentation.

\subsection{Namelist {\tt main}}\texttt{main}
This contains general variables applicable to both static and dynamic
mode.  They are mostly defined in module \texttt{Params}.
\begin{description}
\item[\texttt{tcoul}:] determines whether the Coulomb field should be
  included. Default is {\tt true}.
\item[\texttt{trestart}:] if {\tt true}, restarts the calculation from
  {\tt wffile}. Default is {\tt false}.
\item[\texttt{tfft}:] if {\tt true}, the derivatives of the wave
  functions, but not of the densities, are done directly through FFT.
  Otherwise matrix multiplication is used, but with the matrix also
  obtained from FFT. Default is {\tt true}.
\item[\texttt{mprint}:] control for printer output. If {\tt mprint} is
  greater than zero, more detailed output is produced every {\tt
    mprint} iterations or time steps on standard output.
\item[\texttt{mplot}:] if {\tt mplot} is greater than zero, a printer plot is
  produced and the densities are dumped every {\tt mplot} time steps
  or iterations. Default is~0.
\item{\texttt{mrest}:} if greater than zero, a {\tt wffile} is produced
  every {\tt mrest} iteration or time step. Default is~0.
\item[\texttt{writeselect}]: selects the output of densities by giving
  a string of characters choosing them (see subroutine 
\texttt{write\_densities} for details. Default is {\tt 'r'}, i.~e., only
the density is written.
\item[\texttt{write\_isospin}]: determines whether the densities should
  be output isospin-summed ({\tt false}) or separately for neutrons
  and protons ({\tt true}). Default is {\tt false}.
\item[\texttt{imode}]: selects a static {\tt imode=1} or dynamic {\tt
    imode=2} calculation.
\item[\texttt{nof}]: (number of fragments) selects the initialization.
  {\tt nof=0}: initialization from harmonic oscillator, only for the
  static case; {\tt nof<0}: user-defined initialization by subroutine
  \texttt{init\_user} in module \texttt{User}; {\tt nof>0}:
  initialization from fragment data as determined in {\tt NAMELIST}
  {\tt fragments}.
\end{description}

\subsection{Namelist {\tt grid}}\texttt{grid}
This defines the properties of the numerical grid.
\begin{description}
\item[\texttt{nx}, \texttt{ny}, \texttt{nz}]: number of grid points in the
  three Cartesian directions. They must be even numbers.
\item[\texttt{dx}, \texttt{dy}, \texttt{dz}]: spacing between grid points in fm.
  If only {\tt dx} is given in the input, all three grid spacings
  become equal.
  The grid positions are then set up to be symmetric with the
  coordinate zero centrally between point number {\tt nx/2} and {\tt
    nx/2+1}.
\item[\texttt{periodic}]: chooses a periodic ({\tt true}) or isolated
  ({\tt false}) system.
\end{description}

\subsection{Namelist {\tt static}}\texttt{static}
These input variables control the static calculations.
\begin{description}
\item[\texttt{tdiag}:] if {\tt true}, there is a diagonalization of the
  Hamiltonian during the later (after the 20th) static iterations. This
  20 is hard coded in {\tt static.f90}. Default is {\tt false}.
\item[\texttt{tlarge}:] if {\tt true}, during the diagonalization the new
  wave functions are temporarily written on disk to avoid doubling the
  memory requirements. Default is {\tt false}.
\item[\texttt{nneut}, \texttt{nprot}:] The numbers of neutrons and protons
  in the nucleus. These are used for the harmonic-oscillator
and user initialization.
\item[\texttt{npsi}:] the numbers of neutron ({\tt npsi(1)}) and proton
  ({\tt npsi(2)}) wave functions actually used including unfilled
  orbitals. Again, useful only for harmonic-oscillator or user initialization.
\item[\texttt{radinx}, \texttt{radiny}, \texttt{radinz}:] the radius parameters of the
  harmonic oscillator in the three Cartesian directions, in fm.
\item[\texttt{e0dmp}:] the damping parameter. For its use see subroutine
 \texttt{setdmc}. The default value is 100~MeV.
\item[\texttt{x0dmp}:] parameters controlling the relaxation.  The
  default value is 0.2.  In special cases it may be desirable to
  change this to accelerate convergence.
\item[\texttt{serr}:] this parameter is used for a convergence
  check. If the sum of fluctuations in the single-particle energies,
  \texttt{sumflu} goes below this value, the calculation stops. A
  typical value is {\tt 1.E-5}, but for heavier systems and with
  pairing this may be too demanding.
\end{description}

\subsection{Namelist {\tt dynamic}}\texttt{dynamic}
These are variables controlling the dynamic (TDHF) calculation. 

\begin{description}
\item[\texttt{nt}]: number of time steps to be run.
\item[\texttt{dt}]: the time step in fm/c. A standard value is of the
  order of 0.2 to 0.3~fm/c, it depends somewhat on the value of {\tt
    mxpact}. If the combination of these two is not good enough, the
  calculation becomes unstable after some time, in the sense that the
  norm of the wave functions and the energy drift off and can diverge
 (see Sect.~\ref{sec:accuracy}).
\item[\texttt{mxpact}]: the order of expansion for the exponential
  time-development operator. The predictor (trial) step calculation
  uses {\tt mxpact/2} as the order. For more information see
  Sect.~\ref{sec:accuracy}. 
\item[\texttt{rsep}]: termination condition. If the final state in a
  two-body reaction is also of two-body character, the calculation is
  terminated as soon as the separation distance exceeds {\tt rsep}.
  Units: fm. No default. The purpose of this variable is to prevent
  the calulation of continuing into meaningless configurations, like
  crossing of the boundary.
\item[\texttt{texternal}]: indicates that an external perturbing field is
  used. In this case the namelist {\tt extern} must be
  present. Default: {\tt false}.
\end{description}

\subsection{Namelist {\tt extern}}\texttt{extern}
The variables read here describe the external field that is applied to
get the nucleus into a collective vibration. Details can be found in
the description of module \texttt{External}. It is read only if the
parameter \texttt{texternal} read in namelist \texttt{dynamic} is
true.
\begin{description}
\item[\texttt{ipulse}]: the type of pulse applied. For {\tt ipulse=0} the
  wave function is multiplied with a phase factor that produces an
  initial excitation.For {\tt ipulse=1} a Gaussian time dependence is
  used, for {\tt ipulse=2} a $\cos^2$ one. Default: 0. Details are
  given in Eqs.~(\ref{eq:extgauss}) and (\ref{eq:extcos}).
\item[\texttt{isoext}]: isospin character of the excitation. If this is
  zero, protons and neutrons are exited in the same way. For a value
  of 1, they behave oppositely but with a coupling that leaves the
  center-of-mass invariant. Default: 0.
\item[\texttt{tau0}, \texttt{taut}]: time at which the excitation field
  reaches its maximum, and width of the pulse. No defaults.
\item[\texttt{omega}]: if this is nonzero, the time-dependence of the
  external field gets an additional cosine factor with frequency {\tt
    omega}.
\item[\texttt{radext}, \texttt{widext}]: radius and width of a
  Woods-Saxon-type cutoff factor in radius for the external field.
  Defaults: 100~fm and 1~fm, which practically implies no damping. 
Definition in Eq.~(\ref{eq:extdamp}).
\item[\texttt{amplq0}]: amplitude for quadrupole excitation of the
  $Q_{20}$ type. Defined as usual with respect to the $z$-axis.
\end{description}

\subsection{Namelist {\tt fragments}}\texttt{fragments}
The variables in this namelist control fragment initialization for the
case of \texttt{nof}{\tt >0}. Most quantities are dimensioned for the fragments
and we indicate this by index ``{\tt i}'' in the following.
\begin{description}
\item[\texttt{filename}{\tt (i)}]: the name of the file containing the wave
  functions of fragment {\tt i}.
\item[\texttt{fcent}{\tt (1:3,i)}]: initial position of fragment {\tt i} given
  as three Cartesian coordinate values in fm. The position must be
  such that the complete fragment grid fits inside the new
  computational grid.
\item[\texttt{fix\_boost}]: used only for the two-fragment case. if this
  logical variable is {\tt TRUE}, the initial velocities are
  calculated from the {\tt fboost} values; otherwise from the relative
  motion quantities {\tt ecm}  and {\tt b}.
\item[\texttt{fboost}{\tt (1:3,i)}]: the initial boost of the fragment in the three
  Cartesian directions. It is given as the total kinetic energy in
  each direction in MeV, with the sign indicating positive or negative
  direction. Thus {\tt SUM(ABS(fboost(:,i)))} is the total kinetic
  energy of fragment {\tt i}.
\item[\texttt{ecm}, \texttt{b}]: center-of-mass kinetic energy in MeV and
  impact parameter in fm. Used only if {\tt fix\_boost} is {\tt FALSE}.
  These are the values at infinite distance
  and are corrected using Rutherford trajectories (assuming spherical
  nuclei) for initialization at the finite distance given by
  the {\tt fcent} coordinates. 
\end{description}

\subsection{Namelist {\tt user}}\texttt{user}
This namelist is read only if needed for user initialization (see
module \texttt{User}). Its contents depend on the specific user
initialization and the only thing to be said here is that it should
appear last in the input file. Since the namelist is defined and used
only in module {\tt User}, its name can also be changed arbitrarily,
of course.

\section{Output description}
The code produces a number of output files containing various pieces
of information. 
The bulky observables, such as densities or currents, are selectively
output at certain time steps into special binary output files 
{\em nnnnnn}{\tt .tdd}), where {\em nnnnnn} indicates the iteration or
time step number. These files can then be used for further analysis or
converted to be used as input in visualization codes. Examples of this
are found among the utility codes provided.

The complete set of wave functions is saved at regular
intervals of {\tt mrest} iterations or time steps. Because this leads
to large storage requirements, only the last such file in a run is
kept. It can be used for restarting the calculation or for inputting
fragment wave functions for initializing another calculation.

In {\tt MPI} mode the wave functions are distributed over several
files, each containing only those present on a specific processor. An
additional header file contains the remaining information and can be
used to read the wave functions even on a different processor
configuration. 

Aside from these binary files
there are a number of text files. The
{\tt *.res} files contain one line for each time step or iteration
where output is triggered according to the value of \texttt{mprint}.
There is an explanatory header line in these that has a leading '\#',
so that is treated as a comment by {\tt gnuplot} --- thus {\tt
  gnuplot} can be used immediately to plot the behavior of any one
column of numbers, i.~e., the dependence of a physical quantity on
iteration or time. In addition more complicated output is printed on
standard output, which can be redirected into a file
using shell redirection.

The names of the output files can be adjusted using input variables as
listed in Section~\ref{sec:filenamein}, so that the files are here
denoted by the default names given there. The {\tt *.res} files are
relatively small, so that no mechanism was implemented to suppress
them.

\subsection{File {\tt conver.res}}
This is produced in \texttt{sinfo} only in the static calculation and its
purpose is to give a quick impression of the convergence behavior. The
numbers given in each line are the iteration count, the total energy
in MeV, the relative change in energy from one iteration to the next,
the average uncertainties in the single-particle energies {\tt efluct1} and
{\tt efluct2}, the root-mean-square radius in fm and finally the
deformation parameters $\beta$ and $\gamma$ (see Section
\ref{sec:mulmom}). The latter give an impression as to where the
nuclear shape is ending up.

For the judging of convergence, the {\tt efluct} values are more
important than the change in total energy, since the energy can remain
constant while the wave functions still change considerably.

\subsection{File {\tt monopoles.res}}
At present this file is generated in subroutine
\texttt{moment\_shortprint}, but only in the dynamic mode. It contains
the time, the neutron, proton, and total root-mean-square radii, and
the difference of neutron minus proton root-mean-square radii.

\subsection{File {\tt dipoles.res}}
This file is produced both in the static and dynamic calculations in
subroutines \texttt{sinfo} and \texttt{tinfo}. It contains the iteration or
time step number followed by the three components of the center of
mass vector $\vec{R}$ and those of the difference of proton minus
neutron center-of-mass vectors $\vec{R}^{(T=1)}$, both in fm, for the
definition see Eq.~(\ref{eq:cmcart}). The first of these is useful as
a check to see whether the center of mass drifts off during the
calculation, while the second vector may be useful to look at proton
vs. neutron vibrations.

\subsection{File {\tt quadrupoles.res}}
This is also generated in \texttt{moment\_shortprint} and thus only in
dynamic mode. It contains the Cartesian quadrupole moments
(\ref{eq:Qcart}) for neutrons, protons, and the full mass distribution
followed by the expectation values of $x^2$, $y^2$, and $z^2$ for
neutrons and protons, all in fm$^2$.

\subsection{File {\tt energies.res}}
This is the important monitoring file for the dynamic calculation. It
is written in subroutine \texttt{tinfo} and each line contains the
simulation time, the number of neutrons and protons in the system
(these should be constant, so this is a stability check), the total
energy (again, this should be conserved), the total kinetic energy,
and finally the collective energy {\tt ecoll} separately for neutrons
and protons.  Units for the energies are
all MeV.

\subsection{Standard output}
This contains all the additional information that in most cases is not
needed directly for further processing in, e.~g., graphics programs. If it should
be found necessary to utilize some data from this file, it is in most
cases easy to use {\tt grep} or a scripting language like {\tt PERL}
or {\tt PYTHON} to extract the necessary data. Of course the code can
also be modified to produce additional output files.

The initial part of the output essentially echoes all the data from
the {\tt NAMELIST}s in tabular form, to enable checking the
correctness of input data. In the case of fragment initialization this
is more involved and is discussed below for the dynamic case, since it
is not so common for static calculations.

In general the layout of the information is compact with sequences of
``*'' characters to provide separation between input groups as some
guidance for the eye.

\subsubsection{Static calculation}
The code first prints the current iteration number. Iteration ``0'' refers
to the state before iterations are started, for the later iteration
numbers, the information refers to the end of the iteration.

The an overview of the various energy contributions is printed: the
first part is similar to what in in file {\tt conver.res}, while a
second list shows the energies calculated from the density functional
and split up for the various contributions.

Next there is a simple printer plot of the density distribution in the
$(x-z)$-plane. This is often quite helpful, since it shows what is
going on in the calculation without the need to start a graphics
program, which requires converting the data first.

Next there is a listing of single-particle states. For each state this
shows its parity, occupation probability {\tt wocc} (which is called
$v^2$, as it is interesting mostly in the pairing case), the energy
fluctuations  \texttt{sp\_efluct1} and \texttt{sp\_efluct2}, the norm, the kinetic and total
energies of the state, and finally the expectation values of the three
components of the orbital and spin angular momentum, respectively.

Finally a summary of some integrated quantities is given, separately for
neutrons, protons, and all nucleons: the particle number, the
root-mean-square radius, quadrupole moment, and the average of the
coordinates squared, followed by the center-of-mass components.

Then iterations continue and only one line is printed for each, as
this may be quite slow and it is important to be able to check
progress while the code is running. After \texttt{mprint} iterations
the detailed information is repeated.

\subsubsection{Dynamic calculation}
After echoing the parameters for the dynamic calculation, the fragment
definitions are given and all the resulting information is printed:
the computed boost values in case of twobody initialization, the
properties of the single-particle states read in, including which
index in the fragment file is transferred to which index in the total
set of wave functions.

In case of an external field, the input data is also echoed in a
detailed form.

The time stepping starts and detailed output is produced every 
\texttt{mprint} steps at the end of the time step. Much of it is similar to
the the static case, so only the differences are pointed out.

Because in the dynamic case the situation can have a general
three-dimensional character, the full information on the quadrupole
tensor (\ref{eq:Qcart}) is printed, separately for the neutron,
proton, and total mass distributions. The three eigenvalues and
associated normalized eigenvectors are given, followed by Cartesian
and polar deformation parameters $a_{0}$, $a_{2}$, $\beta$, and
$\gamma$, as defined in Section~\ref{sec:mulmom}.

The separation of the two fragments and its time derivative is printed
next {\bf and repeated every time step}, as examining these quantities
is meaningful only with more frequent sampling.

The energy information is shortened by omitting the quantities not of
interest in the dynamic case. After the printer plot the results of
the two-body analysis are given (for the meaning of the various
quantities see module {\tt Twobody} in the online documentation. 
The Section on ``collision
kinematics'' shows the mass, charge, position, and kinetic energy of
the two fragments. {\bf It should be kept in mind that the two-body
  analysis is only valid if the reaction plane is the $(x-z)$-plane
  and the results printed may not be useful if the physical situation
  is not of two-body nature but the code does not recognize that.}

The single-particle property list and the integrated quantities are as
in the static case, but the energy fluctuations are omitted.

The next time step is then indicated and the fragment separation data
are printed for every time step until after \texttt{mprint} steps the
full output recurs.

\section{Utilities}
A set of short programs is designed to help with further processing of
output from the code. The currently available set is described here.

Most of these routines contain a loop to input a file name from the
terminal. If it is desired to do this in a loop over a set of files, a
simple trick can be used: generate a list of file names using, e.~g.,

{\tt ls -1 *.tdd > list}

and then execute the program with ``list'' as input, e.~g.,

{\tt ./fileinfo < list}

For {\tt Tdhf2Silo} a script {\tt convert} is provided that handles
this (see below). It can easily be adapted to the other utilities and
must be stored in the same directory as the executable utility program
itself in order for the {\tt dirname} command to work properly.

\subsection{Fileinfo}
This is a short program to print information about binary files
generated by Sky3D. It takes the name of either a {\tt *.tdd} or a
wave function file as input and prints out essentially all the
information contained in the header. It can be compiled simply by
executing

{\tt gfortran -o fileinfo fileinfo.f90}

\subsection{Inertia}
This is intended as an example of an analysis code reading {\tt *.tdd}
files and doing some computation, which can be used as a model for
doing similar things. It illustrates looking for the desired field in
the file and taking into account whether it is stored as a total
density or isospin-separated. Being given a filename as input, it
reads the density and calculates the inertia tensor to print all
its 9~components. Compile it using

{\tt gfortran -o Inertia Inertia.f90}

\subsection{Cuts}
This utility reads the density from a file {\em nnnnnn}{\tt .tdd} file and produces
output files named {\em nnnnnn}{\tt rxy.tdd}, {\em nnnnnn}{\tt
  rxz.tdd}, and {\em nnnnnn}{\tt ryz.tdd}, which contains
two-dimensional cuts through the system in the $(x,y)$, $(x,z)$, and
$(y,z)$ plane, respectively. The cuts are evaluated at the origin for
the third coordinate by averaging the two neighboring planes.

These data files are written in such a format that they can be read
by {\tt gnuplot} for use in its commands for 2-dimensional plotting.

This program is intended again as a template that can be modified for
other applications.

\subsection{Overlap}
This is a code to calculate the overlap of two Slater determinants.
Given the names of two wave function files (which must contain
compatible data: dimensions, force, etc.) it reads the wave functions,
generates the matrices of overlaps between one set and the other
separately for neutrons and protons, and then calculates the
determinant of each, which is the overlap between the two Slater
determinants. It prints some summary information: distance between the
centers of mass of each set, minimum and maximum diagonal elements,
maximum absolute value of off-diagonal elements, and finally the
overlaps for protons and neutrons as well as their product.

This code uses subroutines from {\tt LINPACK} (stored at {\tt
  NETLIB.ORG}), which are included in the file {\tt det.f} with
appropriate copyright. It can be compiled using

{\tt gfortran -o overlap overlap.f90 det.f}.

\subsection{Tdhf2Silo}
\label{sec:silo}
This program is quite complicated. It reads a set of {\tt *.tdd} files
and converts them into {\tt Silo} files. {\tt Silo} is a library for
handling scientific datasets developed at Lawrence Livermore National
Laboratory ({\tt https://wci.llnl.gov/codes/silo/index.html}). This is
the most appropriate library to use in conjunction with the LLNL
graphics visualization tool {\tt VisIt} ({\tt
  https://wci.llnl.gov/codes/visit/home.html}), which was found to be
highly suitable for plotting {\tt Sky3D} results and producing
movies. The conversion code is quite flexible in that it decides what
to produce for the different field types: isospin-summed or not,
vector or scalar. They are given appropriate names for {\tt Silo} with
suffixes p and n for protons and neutrons, and x, y, z for the vector
components. In the case of vector fields a variable containing the
vector definition is also written so that the field can be plotted
immediately as a vector field in {\tt VisIt}.

If the user wants another dataset handling method, the code should be
readily adaptable to other libraries. {\tt VisIt} itself has many ways
of importing data, but of course there also alternative 3D
visualization systems.

\section{Running the code}
\subsection{Compilation and linking}
To produce executable files the code comes with several {\tt
  Makefile}s. The standard {\tt Makefile} produces a sequential code
{\tt sky3d.seq}, the file {\tt Makefile.openmp} a parallel code using
{\tt OpenMP}, while {\tt Makefile.mpi} should produce an {\tt MPI}
distributed system code.

The {\tt Makefile}s are written for the {\tt gfortran} compiler and
the commands and options must be adapted if other compilers are used.
{\bf The user may also have to modify the library names and execution
  of the code under MPI will require consulting the local
  documentation or system administrator.}

No attempt was made to select the compiler and linker options
optimally for speed, since experience has shown that optimization at
the cutting edge is highly time-dependent. Thus users should do some
speed tests before embarking on major calculations.

\subsection{External libraries needed}
The {\tt LAPACK} library is used in the code to supply the routines
{\tt ZHBEVD} and {\tt DSYEV}. {\tt LAPACK} should be installed in most
scientific computing centers; if not, the files can be obtained from
{\tt www.netlib.org} and just be added as additional source files to
the code. Note that a complete set of routines called by these two
subroutines must be downloaded.

The other external routine library that is used is {\tt FFTW3}. Again,
it will be preinstalled in most systems. If not, there are two
possibilities:
\begin{enumerate}
\item Download the source code from {\tt www.fftw.org} and compile the
  library yourself. In our experience this worked smoothly. The
  generated library can be installed in a system library directory or
  kept in a user account. In the latter case the use of {\tt -lfftw3}
  in the makefiles does not work anymore and the full path name of the
  library file must be given.
\item Replace it by another FFT routine. This requires quite a bit of
  work: {\tt FFTW} organizes its calls around ``plans'', which
  describe a set of operations to be done on the three-dimensional
  arrays. In \texttt{init\_fft} quite sophisticated plans are set up to,
  for example, transform in the y-direction for all x- and
  z-values.This means that all calls to subroutines beginning with
  {\tt dfftw} have to be examined and possibly replaced by loops over
  one-dimensional FFT transforms. This should be relatively
  straightforward, but there are two more important points to
  consider: 1)~normalization differs between FFT codes. For {\tt FFTW}
  transformation followed by inverse transformation multiplies the
  original data by {\tt nx*ny*nz} and this factor is taken into
  account in several places. 2)~For the non-periodic case the Fourier
  transform in the Coulomb solver uses doubled dimensions in all three
  directions. Some FFT codes have an initialization that sets up the
  transformation factors depending on the dimension; in such cases
  the initialization may have to be repeated.
\end{enumerate}

\subsection{Running with {\tt OPENMP}}
The {\tt OPENMP} version can be compiled using the file {\tt
  Makefile.openmp}, which produces an executable {\tt sky3d.omp}. The
main difference to the sequential makefile is the addition of an
openmp compiler option. Since this depends on the compiler used, it
may have to be modified. For {\tt gfortran} the option is {\tt
  -fopenmp}, while for Intel Fortran it is simply {\tt -openmp}.

For controlling the running of the code the user should set the
environment variable {\tt OMP\_NUM\_THREADS} to the number of parallel
threads to be used (usually the number of processors). In addition it
may be necessary to set {\tt OMP\_STACKSIZE}. The two parallel loops
in {\tt dynamichf} need to store all the density fields in parallel,
and the second loop adds {\tt ps4} to that. Taking into account vector
fields and isospin, a total of~24 threedimensional {\tt COMPLEX(8)}
fields need to be stored, amounting to {\tt nx*ny*nz*24*16} bytes,
which is the stack size needed.

\subsection{Running under {\tt MPI}}
The situation for {\tt MPI} is a bit more complex than for {\tt
  OPENMP}, so that the file {\tt Makefile.mpi} will almost certainly
have to be modified. One crucial difference to the other makefiles is
that the module {\tt Parallel} is now generated from the source file
{\tt parallel.f90}. In addition the compilation commands have to be
adapted; something like {\tt mpif90} will be needed but is
installation dependent. In addition a command like {\tt mpirun} will
be needed for execution; the user is advised to consult local
documentation.

\subsection{Required input}
Here it is just summarized what input is needed for a static or dynamic
calculation. A full description can be found with the documentation
for the {\tt NAMELIST}s.

\subsubsection{Static calculation}

The {\tt NAMELIST}s needed are, in that order:
\begin{quote}
\texttt{files}, \texttt{force}, \texttt{main}, \texttt{grid},
\texttt{static}. 
In addition, if initialization is from fragments, \texttt{fragments},
and for user initialization possibly \texttt{user} (only if the user
initialization requires input).
\end{quote}

\subsubsection{Dynamic calculation}
The {\tt NAMELIST}s needed are, in that order:
\begin{quote}
\texttt{files}, \texttt{force}, \texttt{main}, \texttt{grid},
\texttt{dynamic}, For external field excitation \texttt{extern}, and
in all cases \texttt{fragments}.
\end{quote}

\subsection{Test cases}
To allow checking the proper behavior of the code, we provide three
test cases exercising different functions: a static calculation for
the ground state of $^{16}$O, a dynamic calculation using an external
excitation to stimulate a giant resonance in $^{16}$O, and finally a
sample deep-inelastic collision of two $^{16}$O nuclei. The test cases
directory contains a more detailed description of these cases and what
to look for principally in the results.

\section{Caveats concerning the code}
The user should be aware of the limitations of the code in various
respects but also note some less straightforward procedures for
improving accuracy in certain cases.

\subsection{Static calculations}
Since the main use of the code is expected to be in time-dependent
calculations, the static part is less highly developed. The omission
of all symmetry restrictions, while very useful for innovative
applications with time-dependence, can cause some problems in the
static case.
\begin{enumerate}
\item The spin is not aligned along a fixed direction. Since for
  even-even nuclei Kramers degeneracy operates, two degenerate levels
  will mix in an uncontrolled way to produce an arbitrary spin
  alignment. This can be remedied by diagonalizing the spin operators
  in such a two-level subspace.
\item The center of mass may move away from the origin during the
  iterations. This is typically a very small effect and will be
  corrected when the wave functions are placed at a given position in
  the dynamic initialization. The calculation of observables, however,
  should always use coordinates relative to the real center of mass.
\item In very heavy nuclei sometimes even a rotation was seen, as the
  reoccupation of high-lying levels can change the geometric
  orientation. If this is a serious problem, constraints should be
  introduced or another code used for the static calculation.
\item The harmonic oscillator initialization can be quite deficient
  for heavier nuclei. It is planned to develop a Nilsson-model
  alternative; meanwhile in case of problems the use of wave functions
  from axial or symmetry-restricting codes could be implemented by
  generating a wave function file from such results. Since this will
  involve interpolation, a number of static iterations should then be
  performed using the present code to improve stability.
\end{enumerate}

\subsection{Dynamic calculations}
Here it is important what the required accuracy will be. Most
exploratory calculations will not pose high accuracy demands. There
are several possible ways in which accuracy can be improved if needed:
\begin{enumerate}
\item The initial configuration may be improved. If the fragment
  nuclei are not situated at grid points, one should run a number of
  static iterations with each fragment situated alone in the new
  grid. 
\item If the fragments are deformed, the energy estimate from the
  Rutherford trajectory will not be reliable; in this case it is
  recommended to run a number of dynamic iterations, observe the
  change in relative distance, and then correct the boost energies to
  match the correct velocity of relative motion.
\end{enumerate}

\section{Modifying the code}
Since the code was developed with a view for easy modification, in this
Section we give some advice on how to add new things to it and how to
run simulations.

\subsection{Modifying the Skyrme force}
The code comes with a quite large database of Skyrme force
parametrizations together with appropriate pairing parameters built
in. This is certainly useful to avoid mistakes in the input by having
to indicate only the name of the force. Still there will be a need to
add new forces and even forces with a different density functional to
it, for which we suggest three different approaches.

\subsubsection{Direct parameter input}
If a specific parametrization is not expected to be a permanent
addition to the code, for example if one or several parameters are
varied to study the sensitivity of the results to specific parts of
the density functional, the best way is to use the facility for giving
the parameters directly in the input. This is triggered by using some
force name that is not in the database, in which case the routine
expects all parameters to be given in {\tt NAMELIST forces}.

\subsubsection{Expanding the database}
At present the database of forces is contained in the file {\tt
  forces.data} as a long initialization statement. This has the
advantage of readability and avoids having to make a database file
accessible in every directory used for code applications.

So a new force which will be used more permanently can be added simply
by adding the appropriate lines in {\tt forces.data} and increasing
the number assigned to {\tt nforce} accordingly. There is a slight
danger that the total length of the list will exceed the 255 lines
allowed by the Fortran standard; in this case either remove some
outdated forces, remove the separation lines of all stars, or if there
is still a problem, convert the initialization to {\tt DATA}
statements initializing a smaller number of forces in each case.

\subsubsection{Adding new physics to the density functional}
This can of course require a lot more modifications to the code.
Generally speaking, such a new term will appear not only in the
density functional, but also leads to new contributions in the
single-particle Hamiltonian and may require new types of densities and
currents. In addition, it will involve new force parameters. So in
general the following steps will be needed:
\begin{enumerate}
\item {\bf Parameter definition:} The new parameter can be added to
  the general {\tt Force} type definition in {\tt forces.f90}. This is
  the most logical and systematic way, but we recommend it only if the
  new physics is to be there permanently, since in this case the whole
  database has to be updated to give a default value --- probably zero
  --- to the new parameter for each existing force. The alternative is
  to leave this parameter as a separate entity, which can still be a
  module variable in {\tt Forces} and be input using the same
  namelist. Derived parameters should also be calculated here.

\item {\bf New densities and currents:} everything that can be defined
  directly as a sum over the occupied single-particle wave functions
  should be defined and calculated in module {\tt Densities}. It
  should be a module variable and allocated during initialization
  similar to the densities already defined. Then the contributions of
  the different derivatives of the wave function can be accumulated
  separately as is done for the existing contributions.

\item {\bf Calculation of mean-field components:} subroutine {\tt
    skyrme} in module {\tt Meanfield} is the place where the fields
  appearing in the single-particle Hamiltonian are calculated. The
  difference to module {\tt Densities} is that wave functions are not
  involved directly, but only combinations and derivatives of the
  densities and currents need to be evaluated. The new fields should
  be defined and allocated and then calculated in subroutine {\tt
    skyrme}. The handling of derivatives again can be imitated based
  on the existing terms.

\item {\bf Single-particle Hamiltonian:} The additional contribution
  to the single-particle Hamiltonian must be calculated in subroutine
  {\tt hpsi} of module {\tt Meanfield}. Code has to added to calculate
  how the additional terms act on the input wave function {\tt pinn}
  and the result has to be added to the output wave function {\tt
    pout}. Again, for efficiency the spatial derivatives can be
  handled in separate loops.

\item {\bf Contribution to the energy:} subroutine {\tt integ\_energy}
  in module {\tt Energies} must include an additional contribution of
  the new term, which in this case means simply computing the
  expression for the energy functional. Probably it will be useful to
  add this up in some new variable (defined as a module variable), so
  that the contribution of the new term can be printed out together
  with the other contributions in subroutines {\tt sinfo} of module
  {\tt Static} or subroutine {\tt tinfo} of module {\tt Dynamic},
  respectively.

\item {\bf Output of densities:} it may be desirable to write out the
  new densities, currents, or mean-field contributions into the {\tt
    *.tdd} files. To do that, subroutine {\tt write\_densities} in
  module {\tt Inout} must be modified. A new letter to use for {\tt
    writeselect} must be defined and selected in the {\tt SELECT CASE}
  statement, and then depending on whether it is a scalar or a vector
  field, {\tt write\_one\_density} or {\tt write\_vec\_density} is
  called with a descriptive name given to the field.

  The complication that occurs here is that these writing routines
  assume isospin-dependent fields and output either isospin-separate
  or isospin-summed fields depending on the value of {\tt
    write\_isospin}. If the field has no isospin dependence, the
  statements used to output {\tt wcoul} should be imitated.

\subsection{Analyzing the results in new ways}
 The code provides quite a large number of physical observables in
  its output files. For new applications it may be necessary to look
  at additional ones. There are essentially two ways to implement
  this:
  \begin{enumerate}
  \item If the new observable depends only on density and mean-field
    components, the easiest way is to use the {\tt *.tdd} files, where
    if necessary more fields can be output by modifying the subroutine
    \texttt{write\_densities}. As an example for reading the {\tt *.tdd}
    files, we provide a code calculating the tensor of inertia among
    the utilities.
  \item It becomes more complicated if the wave functions have to be
    used. Here one possibility is to add a call to a user-written
    routine at the beginning of the static or dynamic loops, which
    then can use the array \texttt{psi} in any way desired. This may
    not be a good option if the calculations are lengthy and the
    analysis routine may have to be modified several times. In this
    case it is better to generate a new file name for the {\tt wffile}
    each time \texttt{write\_wavefunctions} is called, similar to the
    way it is done in \texttt{write\_densities}, and to do the
    analysis by reading the wave function files.
  \end{enumerate}

\end{enumerate}

\section*{Acknowledgments}
This work was supported by the Bundesministerium f\"ur Bildung und
Forschung under contract No.\ 05P12RFFTG, by the UK STFC under grant
number ST/J000051/1, and by the U.S. Department of Energy under grant
No.  DE-FG02-96ER40975 with Vanderbilt University.

\bibliographystyle{elsarticle-num} 
\bibliography{Sky3D}

\begin{thebibliography}{10}
\expandafter\ifx\csname url\endcsname\relax
  \def\url#1{\texttt{#1}}\fi
\expandafter\ifx\csname urlprefix\endcsname\relax\def\urlprefix{URL }\fi
\expandafter\ifx\csname href\endcsname\relax
  \def\href#1#2{#2} \def\path#1{#1}\fi

\bibitem{Fet71}
A.~L. Fetter, J.~D. Walecke, {Quantum Theory of Many-Particle Systems},
  McGraw-Hill, New York, 1971.

\bibitem{Mar10aB}
J.~Maruhn, P.-G. Reinhard, E.~Suraud, Simple models of many-fermions systems,
  Springer, Berlin, 2010.

\bibitem{Dre90}
R.~M. Dreizler, E.~K.~U. Gross, {Density Functional Theory: An Approach to the
  Quantum Many-Body Problem}, Springer-Verlag, Berlin, 1990.

\bibitem{Ben03aR}
M.~Bender, P.-H. Heenen, P.-G. Reinhard,
  \href{http://dx.doi.org/10.1103/RevModPhys.75.121}{Self-consistent mean-field
  models for nuclear structure}, Rev. Mod. Phys. 75 (2003) 121.
\newline\urlprefix\url{http://dx.doi.org/10.1103/RevModPhys.75.121}

\bibitem{Dir30}
P.~Dirac, {Exchange phenomena in the Thomas-Fermi-atom }, Proc. Cambridge
  Philos. Soc. 26 (1930) 376.

\bibitem{Bro71aB}
G.~E. Brown, Unified Theory of Nuclear Models and Forces, 3rd Edition,
  North-Holland, Amsterdam, London, 1971.

\bibitem{Rei03a}
P.-G. Reinhard, E.~Suraud, Introduction to Cluster Dynamics, Wiley, New York,
  2004.

\bibitem{Bon76a}
P.~Bonche, S.~E. Koonin, J.~W. Negele, One--dimensional nuclear dynamics with
  time--dependent hartree--fock approximation, Phys. Rev. C 13 (1976)
  1226--1258.

\bibitem{Neg82aR}
J.~W. Negele, The mean--field theory of nuclear structure and dynamics, Rev.
  Mod. Phys. 54 (1982) 913--1015.

\bibitem{Dav85a}
K.~T.~R. Davies, K.~R.~S. Devi, S.~E. Koonin, M.~R. Strayer, {TDHF}
  calculations of heavy--ion collisions, in: D.~A. Bromley (Ed.), Treatise on
  Heavy--Ion Physics, Vol. 3 Compound System Phenomena, Plenum Press, New York,
  1985, p.~3.

\bibitem{Bai87a}
J.~J. Bai, R.~Y. Cusson, J.~Wu, P.-G. Reinhard, H.~St\"ocker, W.~Greiner, M.~R.
  Strayer, Mean field model for relativistic heavy ion collisions, Z. Phys. A
  326 (1987) 269--277.

\bibitem{Ber95a}
H.~Berghammer, D.~Vretenar, P.~Ring, Computer program for the time-evolution of
  a nuclear system in relativistic mean-field theory, Comp. Phys. Comm. 88
  (1995) 293.

\bibitem{Vre05aR}
D.~Vretenar, A.~Afanasjev, G.~Lalazissis, P.~Ring, Relativistic
  hartree-bogoliubov theory: Static and dynamic aspects of exotic nuclear
  structure, Phys. Rep. 409 (2005) 101.

\bibitem{Sim03a}
C.~Simenel, P.~Chomaz, Phys. Rev. C 68 (2003) 024302.

\bibitem{Mar05a}
J.~Maruhn, P.-G. Reinhard, P.~Stevenson, I.~Stone, M.~Strayer, Phys. Rev. C 71
  (2005) 064328.

\bibitem{Umar05a}
A.~S. Umar, V.~E. Oberacker, Phys. Rev. C 71 (2005) 034314.

\bibitem{Nak05a}
T.~Nakatsukasa, K.~Yabana, Phys. Rev. C 71 (2005) 024301.

\bibitem{Has08a}
Y.~Hashimoto, K.~Nodeki, A numerical method of solving time-dependent
  hartree-fock-bogoliubov equation with gogny interaction, arXiv:0707.3083.

\bibitem{Uma09a}
A.~Umar, V.~Oberacker, Density-constrained time-dependent hartree-fock
  calculation of 16o + 208pb fusion cross-sections, Eur. Phys. J. A 39 (2009)
  243.
\newblock \href {http://dx.doi.org/10.1140/epja/i2008-10712-5}
  {\path{doi:10.1140/epja/i2008-10712-5}}.

\bibitem{Loe12a}
N.~Loebl, A.~S. Umar, J.~A. Maruhn, P.-G. Reinhard, P.~D. Stevenson, V.~E.
  Oberacker,
  \href{http://link.aps.org/doi/10.1103/PhysRevC.86.024608}{Single-particle
  dissipation in a time-dependent hartree-fock approach studied from a
  phase-space perspective}, Phys. Rev. C 86 (2012) 024608.
\newblock \href {http://dx.doi.org/10.1103/PhysRevC.86.024608}
  {\path{doi:10.1103/PhysRevC.86.024608}}.
\newline\urlprefix\url{http://link.aps.org/doi/10.1103/PhysRevC.86.024608}

\bibitem{Simenel}
C.~Simenel, \href{http://dx.doi.org/10.1140/epja/i2012-12152-0}{Nuclear quantum
  many-body dynamics}, The European Physical Journal A 48~(11) (2012) 1--49.
\newblock \href {http://dx.doi.org/10.1140/epja/i2012-12152-0}
  {\path{doi:10.1140/epja/i2012-12152-0}}.
\newline\urlprefix\url{http://dx.doi.org/10.1140/epja/i2012-12152-0}

\bibitem{Erl11a}
J.~Erler, P.~Kl\"upfel, P.-G. Reinhard, Self-consistent nuclear mean-field
  models: example skyrme-hartree-fock, J. Phys. G 38 (2011) 033101.
\newblock \href {http://dx.doi.org/10.1088/0954-3899/38/3/033101}
  {\path{doi:10.1088/0954-3899/38/3/033101}}.

\bibitem{Sla51}
J.~C. Slater, Phys. Rev. 81 (1951) 385.

\bibitem{Klu09a}
P.~Kl\"upfel, P.-G. Reinhard, T.~J. B\"urvenich, J.~A. Maruhn,
  \href{http://link.aps.org/doi/10.1103/PhysRevC.79.034310}{Variations on a
  theme by {S}kyrme}, Phys.Rev. C 79 (2009) 034310,
  http://www.arxiv.org/abs/0804.3385.
\newline\urlprefix\url{http://link.aps.org/doi/10.1103/PhysRevC.79.034310}

\bibitem{Klu08a}
P.~Kl\"upfel, J.~Erler, P.-G. Reinhard, J.~A. Maruhn,
  \href{http://dx.doi.org/10.1140/epja/i2008-10633-3}{Systematics of collective
  correlation energies from self-consistent mean-field calculations}, Eur.
  Phys. J A 37 (2008) 343, http://www.arxiv.org/abs/0804.340.
\newline\urlprefix\url{http://dx.doi.org/10.1140/epja/i2008-10633-3}

\bibitem{Bar82a}
J.~Bartel, P.~Quentin, M.~Brack, C.~Guet, H.-B. H{\aa}kansson, Towards a better
  parametrisation of skyrme forces: A critical study of the {SkM} force, Nucl.
  Phys. A 386 (1982) 79--100.

\bibitem{Eng75a}
Y.~M. Engel, D.~M. Brink, K.~Goeke, S.~J. Krieger, D.~Vautherin, Time-dependent
  hartree--fock theory with skyrme's interaction, Nucl. Phys. A 249 (1975)
  215--238.

\bibitem{Per04a}
E.~Perli\ifmmode~\acute{n}\else \'{n}\fi{}ska, S.~G. Rohozi\ifmmode
  \acu\~te{n}\else \'{n}\fi{}ski, J.~Dobaczewski, W.~Nazarewicz, Local density
  approximation for proton-neutron pairing correlations: For\ malism, Phys.
  Rev. C 69~(1) (2004) 014316.

\bibitem{Pot10a}
K.~J. Pototzky, J.~Erler, P.-G. Reinhard, V.~O. Nesterenko, Properties of odd
  nuclei and the impact of time-odd mean fields: A systematic
  skyrme-hartree-fock analysis, Eur. Phys. J. A 46 (2010) 299.
\newblock \href {http://dx.doi.org/10.1140/epja/i2010-11045-6}
  {\path{doi:10.1140/epja/i2010-11045-6}}.

\bibitem{Vau72a}
D.~Vautherin, D.~M. Brink, Hartree--fock calculations with skyrme's interaction
  {I}. spherical nuclei, Phys. Rev. C 5 (1972) 626.

\bibitem{Rei95a}
P.-G. Reinhard, H.~Flocard, Nuclear effective forces and isotope shifts, Nucl.
  Phys. A 584 (1995) 467--488.

\bibitem{Rut95a}
K.~Rutz, J.~Maruhn, P.-G. Reinhard, W.~Greiner, Fission barriers and asymmetric
  ground states in the relativistic mean field theory, Nucl. Phys. A 590 (1995)
  680, http://www.arxiv.org/abs/nucl-th/9610037.

\bibitem{Gre05aB}
W.~Greiner, J.~A. Maruhn, Nuclear Models, Springer Verlag, New York, 1996.

\bibitem{Rin80aB}
P.~Ring, P.~Schuck, The Nuclear Many-Body Problem, Springer--Verl., New York,
  Heidelberg, Berlin, 1980.

\bibitem{Rei97a}
P.-G. Reinhard, M.~Bender, K.~Rutz, J.~Maruhn, An {HFB} scheme in natural
  orbitals, Z. Phys. A 358 (1997) 277,
  http://www.arxiv.org/abs/nucl-th/9705054.

\bibitem{Ben00a}
M.~Bender, K.~Rutz, P.-G. Reinhard, J.~Maruhn, Pairing gaps from nuclear
  mean--field models, Eur. Phys. J. A 8 (2000) 59,
  http://www.arxiv.org/abs/nucl-th/0005028.

\bibitem{Rei82a}
P.-G. Reinhard, R.~Cusson, A comparative study of {H}artree-{F}ock iteration
  techniques, Nucl. Phys. A 378 (1982) 418.

\bibitem{Blu92a}
V.~Blum, G.~Lauritsch, J.~Maruhn, P.-G. Reinhard, Comparison of
  coordinate-space techniques in nuclear mean-field calculations, J. Comp.
  Phys. 100 (1992) 364.

\bibitem{Pre92aB}
W.~H. Press, S.~A. Teukolsky, W.~T. Vetterling, B.~P. Flannery, Numerical
  Recipes in {C}: {T}he Art of Scientific Computing, 2nd Edition, Cambridge
  University Press, New York, 1992.

\bibitem{Cal97a}
F.~Calvayrac, E.~Suraud, P.-G. Reinhard,
  \href{http://dx.doi.org/10.1006/aphy.1996.5654}{Spectral signals from
  electronic dynamics in sodium clusters}, Ann. Phys. (N.Y.) 255 (1997) 125.
\newline\urlprefix\url{http://dx.doi.org/10.1006/aphy.1996.5654}

\bibitem{Rei07a}
P.-G. Reinhard, L.~Guo, J.~A. Maruhn,
  \href{http://dx.doi.org/10.1140/epja/i2007-10366-9}{Nuclear giant resonances
  and linear response}, Eur. Phys. J. A 32 (2007) 19,
  http://www.arxiv.org/abs/nucl-th/0703044.
\newline\urlprefix\url{http://dx.doi.org/10.1140/epja/i2007-10366-9}

\bibitem{Rei92b}
P.-G. Reinhard, From sum rules to {RPA}: 1. nuclei, Ann. Phys. (Leipzig) 504
  (1992) 632.

\bibitem{Loe11a}
N.~Loebl, J.~Maruhn, P.-G. Reinhard,
  \href{http://link.aps.org/doi/10.1103/PhysRevC.84.034608}{Equilibration in
  the time-dependent hartree-fock approach probed with the wigner distribution
  function}, Phys. Rev. C 84 (2011) 034608.
\newblock \href {http://dx.doi.org/10.1103/PhysRevC.84.034608}
  {\path{doi:10.1103/PhysRevC.84.034608}}.
\newline\urlprefix\url{http://link.aps.org/doi/10.1103/PhysRevC.84.034608}

\bibitem{Bon78a}
P.~Bonche, B.~Grammaticos, S.~E. Koonin, Three-dimensional time-dependent
  hartree-fock calculations of {$^{16}{\rm O} + ^{16}{\rm O}$} and {$^{40}{\rm
  Ca} + ^{40}{\rm Ca}$} fusion cross sections, Phys. Rev. C 17 (1978) 1700.

\bibitem{Guo08a}
L.~Guo, P.-G. Reinhard, J.~A. Maruhn,
  \href{http://link.aps.org/doi/10.1103/PhysRevC.77.041301}{Conservation
  properties in the time-dependent {H}artree {F}ock theory}, Phys. Rev. C 77
  (2008) 041301, http://www.arxiv.org/abs/0804.2127.
\newline\urlprefix\url{http://link.aps.org/doi/10.1103/PhysRevC.77.041301}

\bibitem{Cha08aB}
B.~Chapman, G.~Jost, R.~van~der Pas, Using {OpenMP}, {MIT Press}, Cambridge,
  2008.

\bibitem{none12}
{MPI: A Message-Passing Interface Standard, Version 3.0}, {High Performance
  Computing Center}, Stuttgart, 2012.

\bibitem{Fri05a}
M.~Frigo, S.~G. Johnson, The design and implementation of fftw3, Proc. IEEE 93
  (2005) 216.
\newblock \href {http://dx.doi.org/doi:10.1109/JPROC.2004.840301.}
  {\path{doi:doi:10.1109/JPROC.2004.840301.}}

\end{thebibliography}
\end{document}